\documentclass[12pt]{article}
\usepackage[left=2.5cm, right=2.5cm, top=3cm]{geometry}

\usepackage[colorlinks=true, urlcolor=citecolor, linkcolor=citecolor, citecolor=citecolor]{hyperref}
\pdfoutput=1

\title{Bayesian estimation of a semiparametric recurrent event model with applications to the penetrance estimation of multiple primary cancers in Li-Fraumeni Syndrome}

\author{
\\
\normalsize SEUNG JUN SHIN\\
\textit{\small Department of Statistics, Korea University, Seoul, South Korea}\\
\\
\normalsize JIALU LI\\
\textit{\small Department of Bioinformatics and Computational Biology, }\\
\textit{\small University of Texas MD Anderson Cancer Center, Houston, U.S.A}\\
\\
\normalsize JING NING\\
\textit{\small Department of Biostatistics, University of Texas MD Anderson Cancer Center, }\\
\textit{\small Houston, U.S.A}\\
\\
\normalsize JASMINA BOJADZIEVA, LOUISE C. STRONG\\
\textit{\small Department of Genetics, University of Texas MD Anderson Cancer Center, }\\
\textit{\small Houston, U.S.A}\\
\\
\normalsize WENYI WANG$^\ast$\\
\textit{\small Department of Bioinformatics and Computational Biology, }\\
\textit{\small University of Texas MD Anderson Cancer Center, Houston, U.S.A}\\
\normalsize WWang7@mdanderson.org
}

\date{}

\usepackage{subfigure, graphicx, epsf}
\usepackage{amsmath,natbib,verbatim,booktabs,longtable,multirow,lscape}
\usepackage{enumerate, color, url}
\usepackage{mathtools,ctable}
\usepackage{algorithm}
\usepackage[titletoc,toc,page]{appendix}
\usepackage{xcolor}
\usepackage{tabularx,colortbl}

\numberwithin{equation}{section}

\newcommand{\bI}{\mathbf{I}}
\newcommand{\bF}{\mathbf{F}}
\newcommand{\bG}{\mathbf{G}}
\newcommand{\bH}{\mathbf{H}}

\newcommand{\bX}{\mathbf{X}}
\newcommand{\bT}{\mathbf{T}}

\newcommand{\bg}{\mathbf{g}}

\newcommand{\bt}{\mathbf{t}}
\newcommand{\bx}{\mathbf{x}}

\newcommand{\bh}{\mathbf{h}}
\newcommand{\bzero}{\mathbf{0}}

\newcommand{\btheta}{\boldsymbol{\theta}}

\newcommand{\bbeta}{\boldsymbol{\beta}}
\newcommand{\bxi}{\boldsymbol{\xi}}

\newcommand{\bgamma}{\boldsymbol{\gamma}}

\begin{document}

\maketitle

\label{firstpage}

\begin{center}
\textbf{SUMMARY}
\end{center}

	A common phenomenon in cancer syndromes is for an individual to have multiple primary cancers at different sites during his/her lifetime. Patients with Li-Fraumeni syndrome (LFS), a rare pediatric cancer syndrome mainly caused by germline \textit{TP53} mutations, are known to have a higher probability of developing a second primary cancer than those with other cancer syndromes. In this context, it is desirable to model the development of multiple primary cancers to enable better clinical management of LFS. Here, we propose a Bayesian recurrent event model based on a non-homogeneous Poisson process in order to obtain penetrance estimates for multiple primary cancers related to LFS. We employed a family-wise likelihood that facilitates using genetic information inherited through the family pedigree and properly adjusted for the ascertainment bias that was inevitable in studies of rare diseases by using an inverse probability weighting scheme. We applied the proposed method to data on LFS, using a family cohort collected through pediatric sarcoma patients at MD Anderson Cancer Center from 1944 to 1982. Both internal and external validation studies showed that the proposed model provides reliable penetrance estimates for multiple primary cancers in LFS, which, to the best of our knowledge, have not been reported in the LFS literature.\\


\section{Introduction \label{section intro}}

A second primary cancer develops independently at different sites and involves different histology than the original primary cancer; it is not caused by extension, recurrence or metastasis of the original cancer \citep{hayat2007cancer}. Multiple primary cancers (MPC) is a term for the development of primary cancers more than once in a given patient over the follow-up time. The occurrence of MPC is becoming more common due to advances in cancer treatment and related medical technologies, which enable more people to survive certain cancers. The National Cancer Institute estimated that the US population in 2005 included around 11 million cancer survivors, which was more than triple the number in 1970 \citep{curtis1973new}. Furthermore, surviving a given cancer does not necessarily suggest a decreased risk of developing another cancer. For example, \citet{van2014risk} reported that the risk of developing a second primary cancer among survivors of Hodgkin’s lymphoma is 4.7-fold more than that among the general population. The risk of developing MPC varies by genetic susceptibility factors as well. For example, Li-Fraumeni syndrome (LFS), a rare pediatric disease involving higher risk of developing MPC, is associated with germline mutation in the tumor suppressor gene \textit{TP53} \citep{malkin1990germ, eeles1994germline}.

Penetrance is defined as the probability of actually experiencing clinical symptoms of a particular trait (phenotype) given the status of the genetic variants (genotype) that may cause the trait. 
Penetrance plays a crucial role in many genetic epidemiology studies as it characterizes the association of a germline mutation with disease outcomes \citep{khoury1988penetrance}. For example, penetrance is an essential quantity for disease risk assessment, which involves identifying the at-risk individuals and providing prompt disease prevention strategies. To be more specific, popular risk assessment models often require penetrance estimates as inputs \citep{domchek2003application, chen2007meta}. 

The data that motivated our study is a family cohort of LFS collected through probands with pediatric sarcoma treated at MD Anderson Cancer Center (MDACC) from January 1944 to December 1982 and their extended relatives \citep{strong1987genetic, bondy1992segregation, lustbader1992segregation, hwang2003germline, wu2006joint}.  We use ``proband'' to denote the affected individual who seeks medical assistance, and based on whom the family data are then gathered for inclusion in datasets \citep{bennett2011practical}. In the LFS application, the MPC-specific penetrance is defined as
\begin{equation} \label{eq::mpc}
\Pr\left\{\mbox{($k+1$)th primary cancer diagnosis} \mid \mbox{TP53 mutation status, history of $k$ cancers}\right\}, \quad k = 0, 1, 2, \cdots 
\end{equation}
If an individual currently has no cancer history (i.e., $k = 0$), then the MPC-specific penetrance \eqref{eq::mpc} becomes $\Pr\left\{\mbox{First primary cancer diagnosis} \mid \mbox{TP53 mutation status} \right\}$, which has been estimated previously by \citet{WuStrongShete2010} ignoring multiple primary cancers. It shall therefore lead to more accurate cancer risk assessment in LFS for both cancer survivors and no-cancer-history individuals by utilizing more detailed individual cancer histories with MPC.

Few attempts have been made to account for MPC in penetrance estimation. \citet{wang2010estimating} used Bayes rule to calculate multiple primary melanoma (MPM)-specific penetrance, based on penetrance estimates for TP53 mutation carriers, the ratio of MPM patients among carriers and non-carriers, and the ratio of MPM and patients with single primary melanoma (SPM) among carriers. However, that estimation did not account for age and other factors that may contribute to variations observed in patients with SPM and MPM, and relied on previous population estimates of penetrance and relative risk. 

MPC can naturally be regarded as recurrent events, which have been extensively studied in statistics \citep{cook2007statistical}. However, the MPC-specific penetrance estimation from LFS data is more challenging than estimations that use the conventional recurrent event model due to the following reasons. Most individuals (74\%) in the LFS family data have unknown \textit{TP53} genotypes, 
and the LFS data are collected through high-risk probands, e.g., those diagnosed with pediatric sarcoma at MDACC, resulting in ascertainment biases. 
Such bias is inevitable in the study of rare diseases such as LFS as they require an enrichment of cases to achieve a sufficient sample size. 

\citet{shin2017LFScs} recently investigated both of the aforementioned problems for the LFS data under a competing risk framework to provide a set of cancer-specific penetrance estimates. In particular, they defined the familywise likelihood by averaging the individual likelihoods within the family over the missing genotypes, which is possible since the exact distribution of missing genotypes is available according to the Mendelian law of inheritance. The familywise likelihood can minimize the efficiency loss since the missing genetic information is taken into account in its calculation. They also proposed to use the ascertainment-corrected joint (ACJ) likelihood \citep{IversenChen2005} to correct the ascertainment bias for the LFS data.

In this article, we propose a Bayesian semiparametric recurrent event model based on a non-homogeneous Poisson process (NHPP)\citep[]{brown2005statistical, weinberg2007bayesian, cook2007statistical} in order to reflect the age-dependent and time-varying nature of the cancer occurrence rate in LFS. Our preliminary analysis justifies the NHPP model for the LFS data. We develop what we call the ascertainment-corrected familywise likelihood for the proposed NHPP model and estimate the parameters using a Markov chain Monte Carlo (MCMC) algorithm. Then, we provide a set of MPC-specific penetrances for LFS, which, to the best of our knowledge, have never been reported in the literature.

The rest of this paper is organized as follows. In Section \ref{s::theData}, we introduce the LFS family data that motivate this study. In Section \ref{s::exploratoryData}, we provide an explorative analysis for the data to justify our approach. In Section \ref{s::theModel}, we propose a semiparametric recurrent event model for MPC based on NHPP. In Section \ref{s::estimation}, we describe in detail how to construct the familywise likelihood, including the ascertainment bias  correction. We provide the posterior updating scheme via MCMC in Section \ref{s::postMCMC}. We describe a simulation study in Section \ref{s::simuStudy}. In Section \ref{s::application}, we apply the proposed method to the LFS data and obtain the estimated age-at-onset MPC-specific penetrances. We also carry out both internal and external validation analyses. Our final discussion follows in Section \ref{s::discussion}.

\section{Preliminary Analysis of the LFS Data}  \label{s::theData}
\subsection{LFS Data Summary}
The pediatric sarcoma cohort data from MDACC consists of 189 unrelated families, with 17 of them being \textit{TP53} mutation positive families in which there is at least one \textit{TP53} mutation carrier, and 172 being negative ones with no carrier 
(See Appendix G in Supplementary Materials). The \textit{TP53} status was determined by PCR of \textit{TP53} exonic regions.  Ascertainment is
                  carried out through identification of a proband who
                  has 
                  a diagnosis of pediatric sarcoma and who
                  introduces his/her family into the data
                  collection. After a family was ascertained, family members were contacted regularly and continually recruited into the study over 1944 to 1982. Blood
                  samples from members of the family were collected
                  whenever available. The genetic testing of \textit{TP53} was
                  performed on these blood samples, which constitutes
                  the genotype data. Among a total of 3,706 individuals, 964 of them had \textit{TP53} testing results. The age at the diagnosis of each invasive primary tumor for each individual was recorded. The follow-up periods for each family ranges from 22-62 years starting from the acertainment date of probands. Among 570 individuals with a history of cancer, a total of 52 had been diagnosed with more than one primary cancer (Table \ref{tb::numprimary}). Further details on data collection and germline mutation testing can be found in \citet{hwang2003germline} and \citet{peng2017estimating}.

\subsection{Exploratory analysis} \label{s::exploratoryData} 
We first carry out a preliminary analysis of the LFS data to propose a  model that correctly reflects the nature of the data. For simplicity in this analysis, we ignore the family structure. Let $(\bT, V, G, S)$ be a set of data given for an arbitrary individual. For an individual who experiences $K, K = 0, 1, \cdots $ primary cancers, $\bT = (T_{k}; k = 0, \cdots, K)$ denotes the individual’s age at diagnosis of the $k$th primary cancer, with $T_0 = 0$; $V$ is the censored age at which the individual is lost to follow-up, which is assumed to be independent of all $T_k$'s; $G$ denotes the genotype variable, coded as 1 for germline mutation and 0 for wildtype, with a large number of missing values as shown in Table \ref{tb::numprimary}; and $S$ denotes the individual’s sex, coded as 1 for male and 0 for female.

In the analysis of MPC, a primary objective is to model the time to the next cancer given the current cancer history. We let $W_k = T_k - T_{k-1}$ denote the $k$th gap time between two adjacent primary cancers, where $k = 1,2, \cdots$. In analyzing the serial gap times $W_1, W_2, \cdots$, the censoring time $V$, although independent of $T_k$, can be dependent on $W_k$ when the $W_k$'s are not independent \citep{lin1999nonparametric}. This is often referred to as dependent censoring in the literature. Dependent censoring makes it inappropriate to fit marginal models for the $k$th gap times $W_k (k \geq 2)$. For example, \citet{cook2007statistical} showed that ignoring dependent censoring can lead to underestimation of the survival functions of the second and subsequent gap times. 

To check the dependent censoring, we compute the correlation between $W_1$ and $W_2$ using Kendall's $\tau$. Table \ref{tb::numprimary} shows that values of $W_k (k \ge 3)$ are rarely observed in the LFS data, and we therefore exclude them in the analysis. 
Noting that both $W_1$ and $W_2$ can be censored, we use the inverse probability-of-censoring weighted (IPCW) estimates of Kendall's $\tau$ after adjusting for the induced dependent censoring issue \citep{lakhal2010inverse}. In the analysis, we exclude probands who are the index person for family ascertainment. More details about IPCW estimates of Kendall's $\tau$ can be found in Appendix A (Supplementary Materials). Finally, the estimated IPCW Kendall's $\tau$ = $-0.017$ (jackknife estimation of the standard error =0.005), which indicates a statistically significant, but very weak correlation between the two gap times within individuals. We have further calculated Kendall's $\tau$ within
  subgroups of mutation carriers and noncarriers. Neither of the subgroup $\tau$ estimates was
  significantly different from zero.

We computed Kaplan-Meier estimates of survival functions $S_1(t)= \Pr(W_1 > t)$ and $S_2(t)= \Pr(W_2 > t)$, stratified by genotype (See Appendix G in the Supplementary Materials). The risk set used for calculating $S_2(t)$ considers only patients with a single primary cancer (SPC) and MPC starting from the first cancer, while $S_1(t)$ includes all individuals. For both \textit{TP53} mutation carriers and non-carriers or untested individuals, the lengths of the first and second gap times are not identically distributed, with the first gap time significantly longer than the second one. This suggests a time trend in the process where the rate of event occurrence increases with age. Moreover, the mutation carriers appear to have different length distributions for wildtype and untested individuals. This empirical difference in successive survival also suggests the importance of providing subgroup-specific and MPC-specific penetrance.

\section{Model} \label{s::theModel} 

\subsection{Semiparametric recurrent event model for MPC} \label{s::spReModelMPC} 

Viewing the MPC as recurrent events that occur over time, we employ a counting process to model the MPC. 

Let $\{N(t), t \geq 0 \}$ be the number of primary cancers that an individual experiences by age $t$. The intensity function $\lambda(t|H(t))$ that characterizes the counting process $N(t)$ is defined as
\begin{align}
\lambda(t|H(t)) = \lim_{\triangle t \to 0} \frac{\Pr \{ N(t + \triangle t) -N(t) >0 |H(t) \}} {\triangle t}
\end{align}
where $H(t)$ denotes the event history up to time $t^-$, i.e., $H(t) = \{N(s), 0 \le s < t\}$, with $t^-$ being a time infinitesimally before $t$ \citep{cook2007statistical, ning2015dependence}.


For the LFS data, we incorporate a covariate $\bX(t) = \{G, S, G \times S, D(t), G \times D(t)\}^T$ into the Poisson process model, where $D(t)$ is a time-dependent, but periodically fixed MPC variable that is coded as 1 if $t > T_1$ and 0 otherwise. We propose the following multiplicative model for the conditional intensity function given $\bX(t)$ as 
\begin{align}
\lambda(t|\bX(t), \xi_i) = \xi_i \lambda_0 (t) \exp( \bbeta^T \bX(t)),
\end{align}
where $\bbeta$ denotes the coefficient parameter that controls effect of covariate $\bX(t)$ on the intensity and $\lambda_0(t)$ is a baseline intensity function. 
Here, $\xi_i$ is the $i$th family-specific frailty used to account for the within-family correlation induced by non-genetic factors that are not included in $\bX(t)$. We remark that $\xi_i$ allows us to relax the assumption that the disease histories are conditionally independent given the genotypes. We consider the gamma frailty model that assumes $\xi_1, \cdots, \xi_I \stackrel{iid}{\sim} Gamma(\phi, \phi)$, where $I$ denotes the number of families. The gamma frailty model has been used as a canonical choice \citep{duchateau2007frailty} due to the mathematical convenience. Recalling that $E(\xi_i) = 1$ and $var(\xi_i) = \phi^{-1}$, a large value of $\phi$ indicates that the within-family correlation is negligible, and we can drop the frailty term to obtain a more parsimonious model.


There are several choices for the baseline intensity function. Constant or polynomial baseline intensity can be used due to its simplicity, but it may be too restrictive in practice. As an alternative, the piecewise constant model has been widely used due to its flexibility. However, the selection of knot points may be subjective, and it always produces a non-smooth function estimate, which is not desired in some applications.  We propose to employ Bernstein polynomials to approximate the cumulative baseline intensity function, $\Lambda_0(t) = \int \lambda_0(u) du$, which is monotonically increasing. Bernstein polynomials are widely used in Bayesian nonparametric function estimation with shape constraints. Assuming $t \in [0,1]$ without loss of generality, Bernstein polynomials of degree $M$ for $\Lambda_0(t)$ are $B_M(t;\Lambda_0) = \bgamma^T \bF_M(t)$, where the known parameter vector $\bgamma = (\gamma_1, \cdots, \gamma_M)$ with $\gamma_m = \Lambda_0(\frac{m}{M}) - \Lambda_0(\frac{m-1}{M}), m = 1, \cdots, M$ and $\Lambda_0(0) = 0$; and $\bF_M(t) = (F_M(t, 1), \cdots, F_M(t, M))^T$, with $F_M(t, m)$ being the beta distribution function with parameters $m$ and $M-m+1$ evaluated at $t$ \citep{mckay2011variable}. We restrict $\gamma_m \ge 0, m = 1, \cdots, M$ to have $\Lambda_0$ monotonically increasing. The Bernstein-polynomial model for $\lambda_0(t)$ is then obtained by 
\begin{align}
\lambda_0 (t) \approx \frac{d}{d t} B_M(t;\Lambda_0) = \bgamma^T {\mathbf{f}}_M(t),
\end{align}
where ${\mathbf{f}}_M(t)^T = (f_M(t, 1), \cdots, f_M(t, M)))$ denotes the beta density with parameters $m$ and $M-m+1$ evaluated at $t$.

A large value of $M$ provides more flexibility to model the shape of baseline rate function, but at the cost of increased computations. \citet{GelfandMallick1995} empirically showed that a relatively small value of $M$ works well in practice, and we assume $M = 5$ in the upcoming analyses.

Finally, the proposed semi-parametric model for the intensity function of NHPP is given by
\begin{align*}
\lambda(t|\bX(t), \xi_i) = \xi_i \bgamma^T \mathbf{f}_M(t) \exp \{\bbeta^T \bX(t)\}.
\end{align*}

\subsection{MPC-specific penetrance} \label{ss::penetranceCalculation}
The MPC-specific age-at-onset penetrance defined in \eqref{eq::mpc} is equivalently rewritten as 
\begin{align} \label{eq::penetrance}
\Pr (W_{k+1} \le w |T_k, \bX(t)),
\end{align}
which is identical to $\Pr (W_{k+1} \le w |T_k, \bX(T_k))$ since $D(t)$ is periodically fixed. The MPC-specific penetrance \eqref{eq::penetrance} is then obtained by marginalizing out the random frailty $\xi$ as follows:
\begin{align*}
\Pr (W_{k+1} \le w | T_{k}= t_{k}, \bX(t_k) ) & = 1 - \int_{0}^{\infty} \exp \left( - \int_{t_{k}}^{t_{k}+w} \lambda(u|\bX(u), \xi) du \right) f(\xi|\phi) d\xi  \\
&= 1- \Bigg( \frac{\phi}{\phi + \int_{t_{k}}^{t_{k}+w} \lambda(u|\bX(u)) du} \Bigg)^{\phi},
\end{align*}
where $f(\xi|\phi)$ is the gamma density function of the frailty $\xi$ given $\phi$, and $\lambda(t|\bX(t)) = \lambda_0 (t)\exp\{\bbeta^T \bX(t)\}$.

\section{Computing Likelihood} \label{s::estimation} 
In this work, the computing likelihood is not trivial due to a large number of missing genotypes and the ascertainment bias. In this section, we propose an ascertainment-bias-corrected familywise likelihood to tackle these issues.

Let $v_{ij}$ and $K_{ij}$ respectively denote the censoring time and the total number of primary cancers developed for individual $j = 1, \cdots, n_i$ from family $i = 1, \cdots, I$. Suppose we are given a set of data $(\bt_{ij}, v_{ij}, g_{ij}, s_{ij})$, where ${\bt_{ij} =  \{ t_{ij,k}: k = 1, \cdots, K_{ij}\}^T }$ and $g_{ij}$ and $s_{ij}$ are the observed genotype and sex, respectively. 
Given the data, we can easily define the observed version of $D(t)$ denoted by $d_{ij}(t)$ as 1 if $t > t_{ij,1}$ and 0 otherwise, and $\bx_{ij}(t) = \{g_{ij}, s_{ij}, g_{ij} \times s_{ij}, d_{ij}(t), g_{ij} \times d_{ij}(t) \}^T$.

\subsection{Individual likelihood}

Let $t_{ij,0} = 0$ and $v_{ij} \geq  t_{ij,K_{ij}}$, conditioning on $\xi_i$, the likelihood contribution of the $k$th event since the $(k-1)$th event is
\begin{align}
\lambda(t_{ij,k}|\bx_{ij}(t_{ij,k}), \xi_i) \exp \bigg( - \int_{t_{ij,k-1}}^{t_{ij,k}} \lambda(u|\bx_{ij}(u), \xi_i) du \bigg),
\end{align}
\citep{cook2007statistical}. 

Let $\bx_{ij} = \{ \bx_{ij}(t_{ij,k}), k = 1, \cdots, K_{ij} \}$ and $\btheta = (\bbeta, \bgamma, \phi)$ denote the parameter vectors of interest. Given $\xi_i$, the likelihood of the $j$th individual of the $i$th family with primary cancer events $\bh_{ij} = (\bt_{ij}, v_{ij})$, denoted by $\Pr(\bh_{ij}| \bx_{ij}, \btheta, \xi_i)$ is 
\begin{align} \label{eq::individual.likelihood}
\Pr(\bh_{ij}| \bx_{ij}, \btheta, \xi_i) 
\propto &  
\left\{ {\prod_{k=1}^{K_{ij}} \lambda(t_{ij,k} | \bx_{ij}(t_{ij,k}), \xi_i)  } \right\} \times \nonumber \\
& \exp \left\{ - \sum_{k=1}^{K_{ij}} \int_{t_{ij,k-1}}^{t_{ij,k}} \lambda(u| \bx_{ij}(u), \xi_i) du \right\} \times
\exp \left\{ - \int_{t_{ij, K_{ij}}}^{v_{ij}} \lambda(u|\bx_{ij}(u), \xi_i) du \right\}.
\end{align}

\subsection{Familywise likelihood} \label{ss::family.like}

Tentatively assuming that the covariates $\bx_{ij}$ are completely observed for every individual, the likelihood for the $i$th family is simply given by $\prod_{j=1}^{n_i} \Pr(\bh_{ij}| \bx_{ij}, \btheta, \xi_i)$. However, in the LFS data, most individuals have not undergone testing for their \textit {TP53} mutation status. For simplicity, we partition the covariate vector $\bx_{ij} = \{\bg_{ij}, \bg_{ij}^c\}$, where $\bg_{ij}$ and $\bg_{ij}^c$ denote the covariates that are related and unrelated to the genotype $g_{ij}$, respectively. Let $\bh_{i} = (\bh_{i1}, \cdots, \bh_{in_i})$, $\bg_i = (\bg_{i1}, \cdots, \bg_{in_i})$ and $\bg_i^c = (\bg_{i1}^c, \cdots, \bg_{in_i}^c)$. Due to a large number of family members without genotype information, $\bg_i$, we further introduce $\bg_{i,obs}$ and $\bg_{i,mis}$ to respectively denote the observed and missing parts of genotype vector $\bg_i$, i.e., $\bg_i = \{\bg_{i,obs}, \bg_{i,mis}\}$. The familywise likelihood for the $i$th family is naturally defined by 
\begin{align} \label{eq:flike}
\Pr(\bh_{i}| \bg_{i, obs}, \bg_{i}^c, \btheta, \xi_i),
\end{align}
while its evaluation is not trivial since $\bh_{i1}, \cdots, \bh_{in_i}$ are correlated through $\bg_{i, mis}$. 

To tackle this issue, we employ Elston-Stewart's peeling algorithm to recursively calculate \eqref{eq:flike} \citep{elston1971general, lange1975extensions, FernandoStrickerElston1993}. Let us suppress the conditional arguments in \eqref{eq:flike} except $\bg_{i,obs}$ for simplicity. The peeling algorithm is developed to evaluate the pedigree likelihood $\Pr(\bh_i)$, not $\Pr(\bh_i | \bg_{i, obs})$, accounting for the probability distribution of genotype configurations of all family members (e.g., $3^n$ genotype configurations for one gene and $n$ family members). It proceeds by recursively partitioning a large family into smaller ones. An illustrative example of the peeling algorithm for the familywise likelihood evaluation is given in Appendix B (Supplementary Materials). Notice that if there is no genotype observed, i.e., $\bg_i = \bg_{i, mis}$, then \eqref{eq:flike} can be evaluated by directly applying the peeling algorithm. We have made slight modification on the peeling algorithm to include known genotype information of some family members in our data \citep{shin2017LFScs}.

\subsection{Ascertainment bias correction} \label{ss::ascertainment.bias}
Ascertainment bias is inevitable in studies of rare diseases like LFS because the datasets are usually collected from a high-risk population. For example, our LFS dataset is ascertained through proabands diagnosed at LFS primary cancers such as pediatric sarcoma at MD Anderson Cancer Center, and therefore has oversampled LFS primary cancer patients. Such ascertainment must be properly adjusted to generalize the corresponding results to the population, for which the familywise likelihood \eqref{eq:flike} alone is not sufficient.

We propose to use an ascertainment-corrected joint (ACJ) likelihood \citep{KraftThomas2000,IversenChen2005}. Introducing an ascertainment indicator variable ${\cal A}_i=1$ that takes 1 if the $i$th family is ascertained and 0  otherwise, the ACJ likelihood for the $i$th family is given by  
\begin{align} \label{eq::acj}
	\Pr(\bh_i, \bg_{i, obs}| \bg_{i}^c,\btheta, \xi_i, {\cal A}_i=1) 
	\propto \frac{\Pr(\bh_{i} | \bg_{i,obs},  \bg_{i}^c, \btheta, \xi_i)}{\Pr({\cal A}_i=1 | \bg_{i}^c, \btheta, \xi_i)}.
\end{align}
That is, the ACJ likelihood corrects the ascertainment bias by inverse-probability weighting \eqref{eq:flike} by the corresponding ascertainment probability. Now, the ascertainment probability, the denominator in \eqref{eq::acj}, is the likelihood contribution of the proband, computed as follows:
\begin{align} \label{eq::acs.prob}
  & \Pr( {\cal A}_i=1|\bg_{i}^c,\btheta, \xi_i) \nonumber \\
= & \sum_{g\in\{0,1\}} \left[\lambda(t_{i1,1}|\bx_{i1}(t_{i1,1};g), \xi_i) \exp \left\{ - \int_{0}^{t_{i1,1}} \lambda(u|\bx_{i1}(u), \xi_i) du \right\}\right]\Pr(G_{i1} = g | \bg_{i1}^c),
\end{align}
where $\bx_{i1}(t;g) = \{g, s_{i1}, g \times s_{i1}, d_{i1}(t), g \times d_{i1}(t) \}^T$ for the proband in family $i$. Notice this likelihood is marginalized over genotype since the genotype information for the proband is not available when  the ascertainment decision is made. In general, the covariate specific prevalence $\Pr(G_{i1} = g|\bg_{i1}^c)$ is assumed to be known. In our LFS application, we assume the \textit{TP53} mutation prevalence is independent of all non-genetic variables and therefore the conditional prevalence is equal to the unconditional prevalence $\Pr(G_{i1}=g)$ for a general popoulation, which can be calculated from the mutated allele frequency denoted by $\psi_A$: $\Pr(G_{i1}=0) = (1- \psi_A)^2$ and $\Pr(G_{i1}=1) = 1- (1- \psi_A)^2$. Here $\psi_A = 0.0006$ for \textit{TP53} mutations in the Western population \citep{lalloo2003prediction}.

Finally, the ACJ familywise likelihood for the LFS data is then given by 
\begin{align*}
\Pr(\bH, \bG_{obs} | \bG^c, \btheta, \bxi, \boldsymbol{\cal A} )
\propto 
\prod_{i=1}^I \frac{\Pr(\bh_{i} | \bg_{i,obs},  \bg_{i}^c, \btheta, \xi_i)}{\Pr({\cal A}_i=1 | \bg_{i}^c, \btheta, \xi_i)}
\end{align*}
where $\bH = (\bh_1, \cdots \bh_I)$, $\bG_{obs} = (\bg_{1, obs}, \cdots, \bg_{I, obs})$, $\bG^c = (\bg_{1}^c, \cdots, \bg_{I}^c)$, and $\bxi = (\xi_1, \cdots, \xi_I)$, and $\boldsymbol{\cal A} = ({\cal A}_1, \cdots, {\cal A}_I)$.

\section{Posterior Sampling through MCMC}   \label{s::postMCMC}
We set an independent normal prior for $\boldsymbol \beta$ where $\bbeta \sim N(\bzero, \sigma^2\bI)$ where $\bzero$ and $\bI$ denote a zero vector and an identity matrix, respectively, and $\sigma = 100$ for vague priors. We assign nonnegative flat priors for $\bgamma \sim Gamma(0.01,0.01)$ for the baseline intensity. We assume a gamma prior for $\phi \sim Gamma(0.01, 0.01)$. The joint posterior distribution of $(\btheta, \phi, \bxi)$ is 
\begin{align}
\Pr(\btheta, \phi, \bxi | \bH, \bG_{obx}, \bG^c, \boldsymbol{\cal A} ) \propto 
\Pr(\bH, \bG_{obs} | \bG^c, \btheta, \bxi, \boldsymbol{\cal A} ) \Pr(\btheta) \Pr(\bxi|\phi)  \Pr(\phi).
\end{align} 
where $\Pr(\btheta)$ and $\Pr(\phi)$ denote prior distributions, and $\Pr(\bxi|\phi)$ is a frailty density that we assume to follow gamma distribution. We use a random walk Metropolis-Hastings-within-Gibbs algorithm to generate 100,000 posterior estimates in total, with the first 5,000 as burn-in. Details about the algorithm steps, R code and convergence diagnostics can be found in the Appendix C (Supplementary materials).

\section{Simulation Study} \label{s::simuStudy}
\begin{figure}
\begin{center}
\begin{tabular*}{\columnwidth}{c@{\extracolsep{\fill}} 
                                                 c@{\extracolsep{\fill}}                     
                                                 c@{}}  \hline

\hline
     		& Frailty    & No Frailty          \\  \hline
$\beta_1$   & 0.1960 (0.0586) & 0.1976 (0.0279) \\ 
$\beta_2$   & 0.3398 (0.0421) & 0.4332 (0.3083) \\ 
$\lambda_0$ & 0.0001 (0.0001) & 0.0001 (0.0001) \\
$\phi$      & 1.6360 (1.5135) & {Not Available} \\ \hline
\end{tabular*}

\subfigure{
\includegraphics[width = 0.31\textwidth]{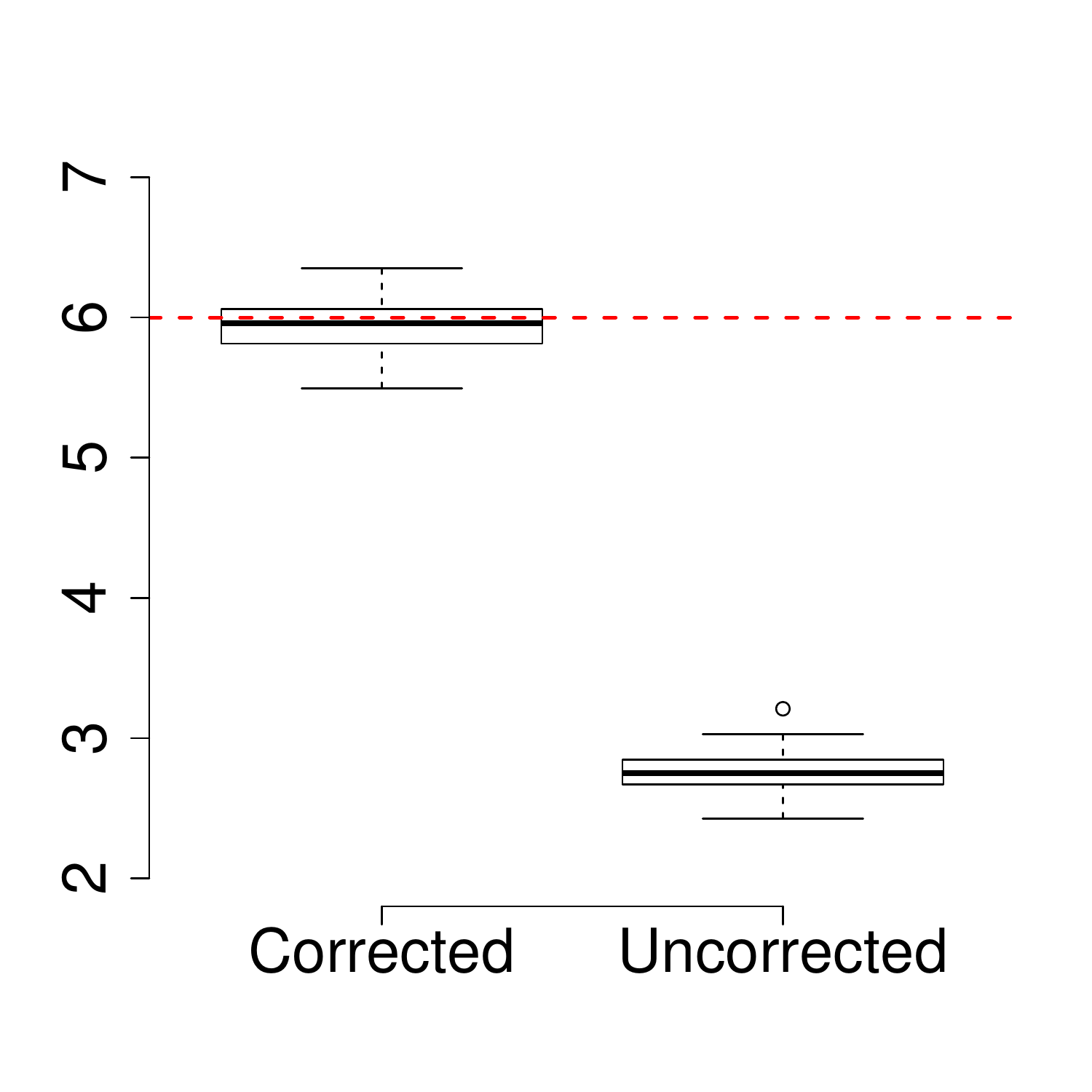}}
\subfigure{
\includegraphics[width = 0.31\textwidth]{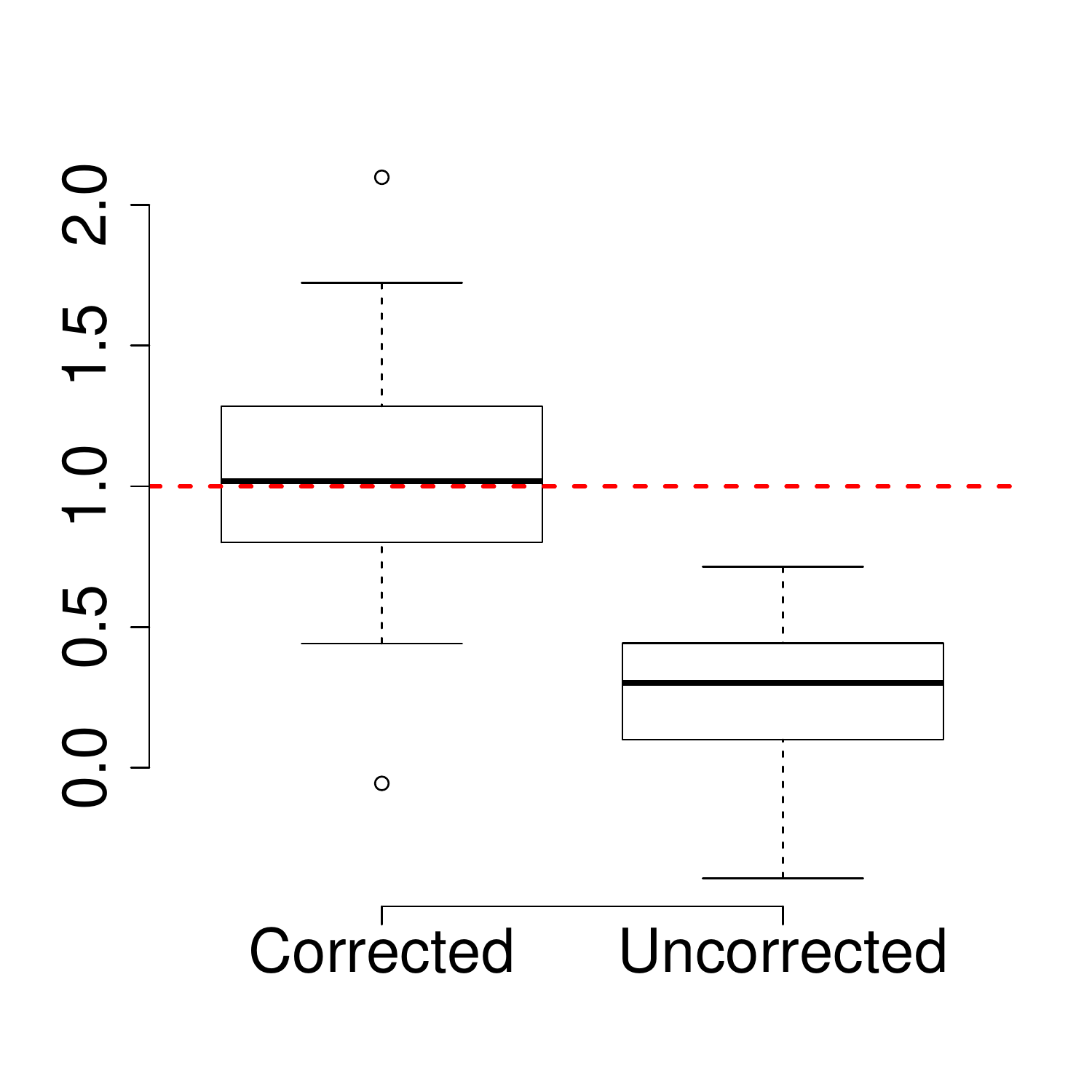}}
\subfigure{
\includegraphics[width = 0.31\textwidth]{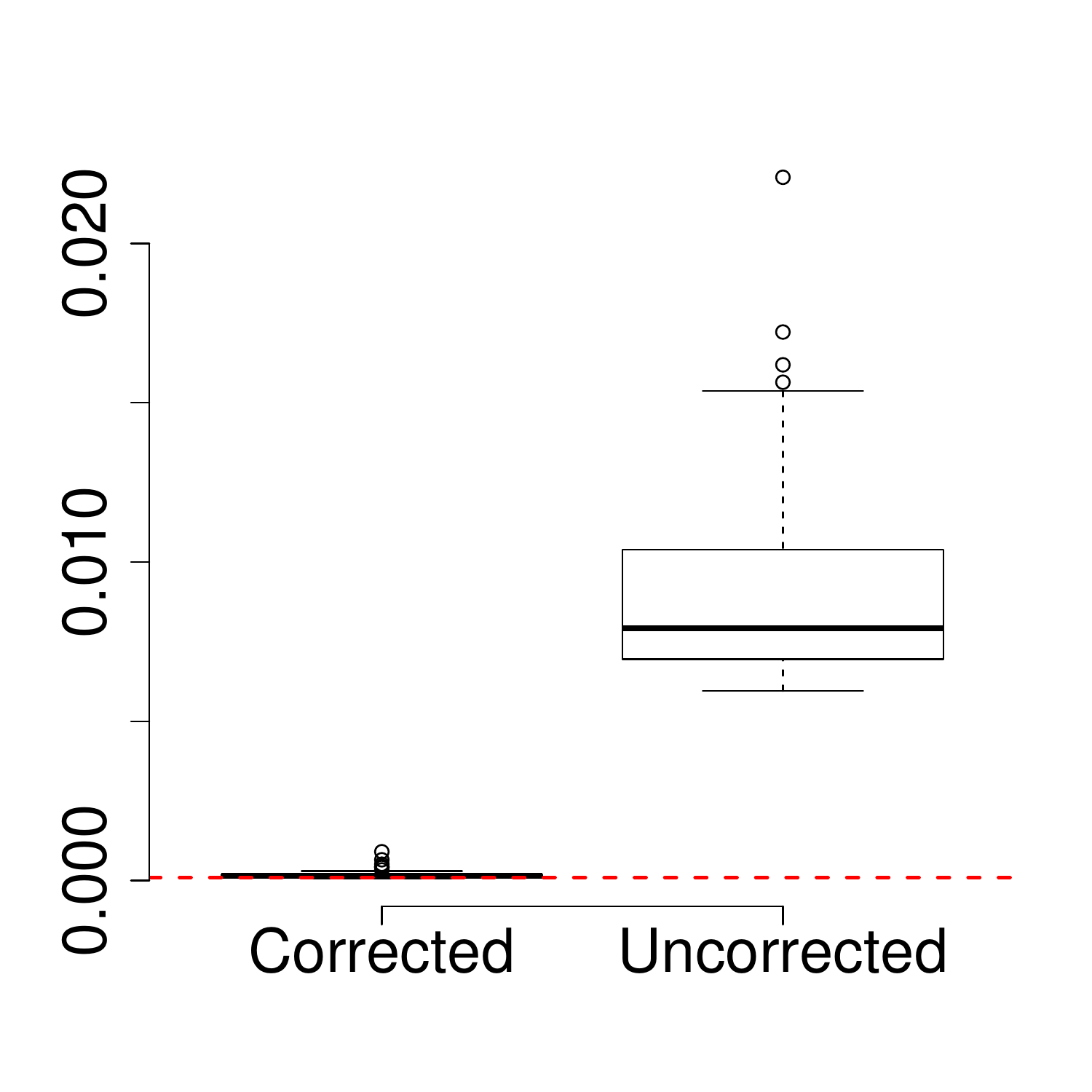}}

\caption{Simulation Results: The top panel shows boxplots of the estimates from 100 independent with the corresponding true values depicted in red dashed line. The Bottom panel reports root mean squared errors and absolute biases (in parenthesis) of the
  estimates from the models with and without frailty term. The truths for the simulation data are: $\beta_1=6$, $\beta_2=1$, $\lambda_0=0.0005$} \label{fg::sim}
\end{center}
\end{figure}

We simulated family history data containing patients with single and multiple primary cancers as follows:
\begin{enumerate}
\item We first simulated the genotype of the proband by $G \sim Bernoulli(0.001)$, based on which we generated the first and second gap times, $W_1$ and $W_2$, from the exponential distribution with the rate parameter being
\begin{align}
\lambda(t|G, D) = \xi \lambda_0 (t) \exp( \beta_1 G + \beta_2 D),
\end{align}
where $D= 0$ for $W_1$ and $D = 1$ for $W_2$. We set a constant baseline $\lambda_0 (t)=0.0005$, and $\beta_1=6$ and $\beta_2=1$. We simulated $\xi \sim Gamma(\phi,\phi)$ with $\phi = 1$. The two gap times were then compared to censoring time generated by $V \sim Exponential(0.5)$ to determine the event indicator. To mimic the ascertainment procedure, for the family data simulation, we retained only the probands with at least one primary cancer observed.
\item Given the proband's data, we generated his/her family data for three generations. The complete pedigree structure for the simulation is depicted in Appendix G (Supplementary materials). We set the genotype of all family members as $G=0$ if the proband was a non-carrier. If the proband was a \textit{TP53} mutation carrier, one of his/her parents was randomly set as a carrier, and the proband's siblings and offsprings were set independently as carriers with a probability of 0.5. The offspring   of the proband's siblings were also randomly set as carriers with a probability of 0.5 if the proband's siblings were carriers. To mimic the scenario of the rate of a rare mutation such as that of the \textit{TP53} gene, all family members with non-blood relationships with the proband were set as non-carriers. 
\item We simulated the first two gap times and the cancer event indicators for the probands' relatives as we did for the probands. We simulated a total of 100 such families, each of which had 30 family members. To mimic the real scenario in which genotype data are not available for a majority of family members, we randomly removed 70\% of the genotype information from non-proband family members.
\end{enumerate}
We applied the proposed methods to the simulated data. We generated 5000 posterior samples from the MCMC algorithm, with the first 1000 as burn-in, and checked that the MCMC chains converged well. Simulation results based on 100 independent repetitions are summarized in Figure \ref{fg::sim}. The proposed method can successfully recover the true parameters. See the top panel for the comparison to a model without frailty, in terms of the root mean squared error and absolute bias, and the bottom panel for the comparison to a model without ascertainment bias correction.

\section{Case Study} \label{s::application} 
We applied our method to the LFS data (Section \ref{s::exploratoryData}) and estimated the parameters using the MCMC algorithm as described in Section \ref{s::postMCMC}. We performed cross-validation, in which we compared our prediction of the 5-year risk of developing the next cancer given the individual's cancer history and genotype information with the observed outcome, based on our penetrance estimates. We also compared our penetrance results with population estimates and the results in previous studies on \textit{TP53} penetrance. 

\subsection{Model fitting}
We fit our model to the LFS data up to the second cancer event due to the limited number of individuals with three or more primary cancers in the dataset (Table \ref{tb::numprimary}). Our model contains three relevant covariates: genotype, sex, and cancer status at time $t$, respectively denoted by $G$, $S$, and $D(t)$. We also included two interaction effects on genotype. 

We applied the proposed method to the entire dataset to obtain penetrance estimates for SPCs and MPCs given the \textit{TP53} mutation status. We first conducted a sensitivity analysis, which showed that the penetrance estimates are not overly sensitive to the choice of priors. The results of the sensitivity analysis are provided in Appendix D (Supplementary materials). 

We then computed the deviance information criterion (DIC) to identify the best set of covariates. We compare five different combinations of $G, S$, and $D(t)$. We observe that the simplest model with $\{G, S, D(t)\}$ achieves the minimum DIC value. However, we decided to select the second best model in terms of the DIC, with $\{G, S, D(t), G \times S, G \times D(t)\}$ as our final model since it has been reported that cancer status has different effects on cancer risk for mutation carriers and non-carriers \citep{van2014risk, mai2016risks}. All posterior estimates of the model generated from the MCMC algorithm converged well and had reasonable acceptance ratios. See Appendix E (Supplementary materials) for the model comparison results. 

Table \ref{tb::posteriorEstimates} contains summaries of the posterior estimates for both the frailty and no-frailty models. We observe that the variance of frailty, $\phi^{-1}$, is estimated to be quite small, which indicates that the no-frailty model may be preferred. It turns out that both models produce nearly identical penetrance estimates (see Appendix E in the Supplementary Materials), and we decide to use the no-frailty model to analyze the LFS data. The genotype has dominant effects on increasing cancer risk, both through a main effect and through interaction with the cancer history, as expected from the exploratory analysis (Section \ref{s::exploratoryData}).

\begin{figure}
	\centering
	\begin{subfigure}[]
		\centering
		\includegraphics[width=0.44\linewidth]{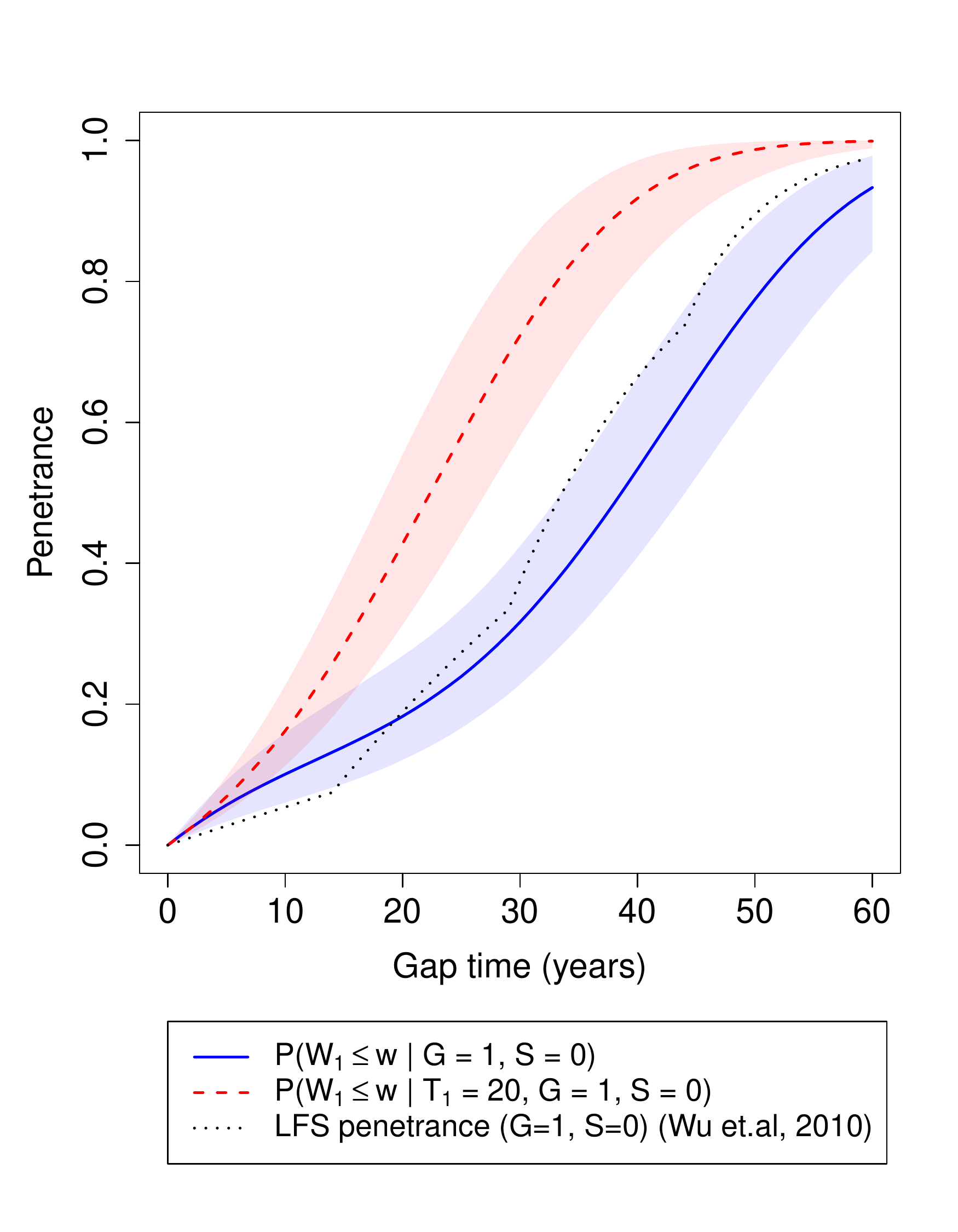}
		\label{fig:a}
	\end{subfigure} 
	\begin{subfigure}[]
		\centering
		\includegraphics[width=0.44\linewidth]{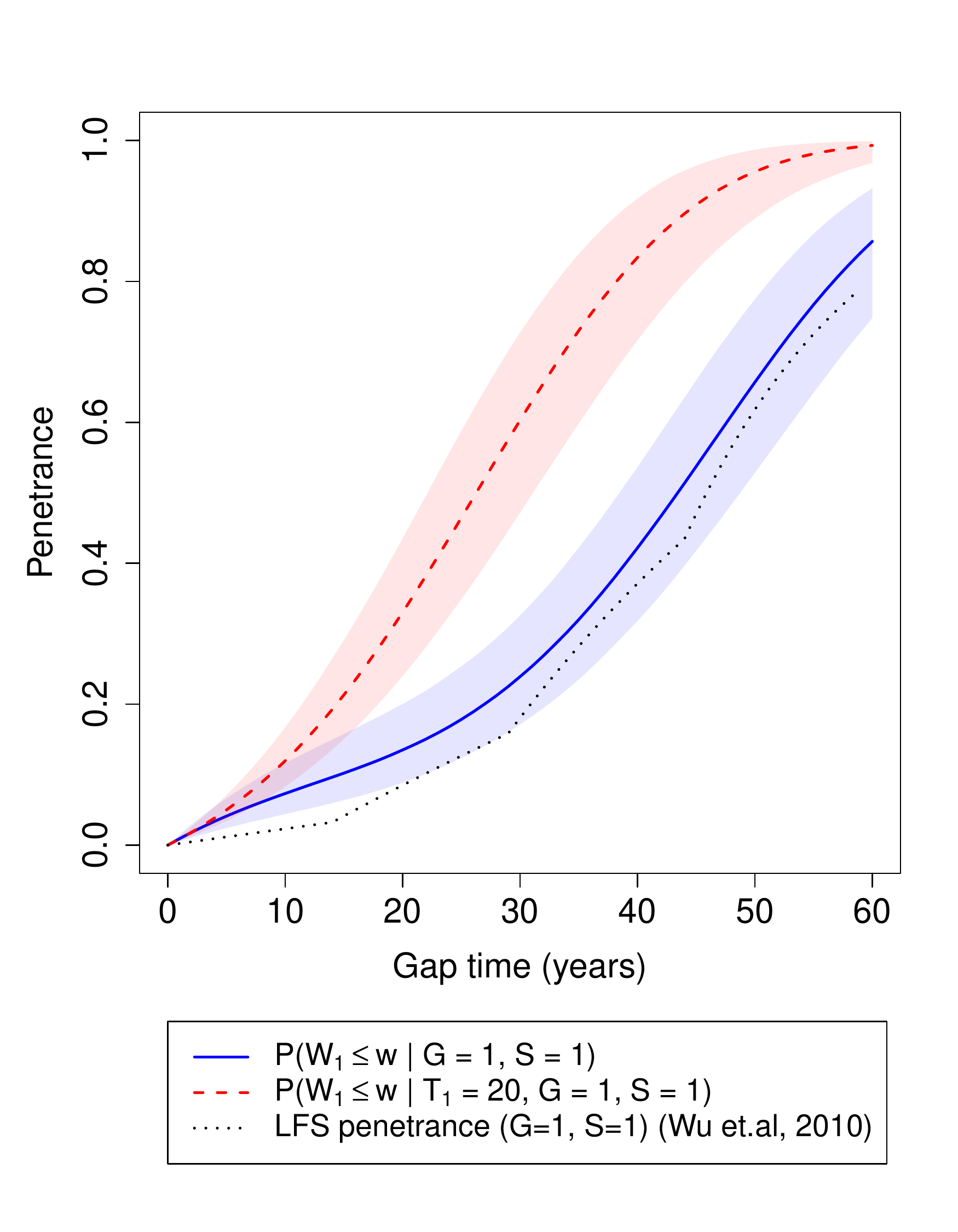}
		\label{fig:b}
	\end{subfigure}
	\begin{subfigure}[]
		\centering
		\includegraphics[width=0.44\linewidth]{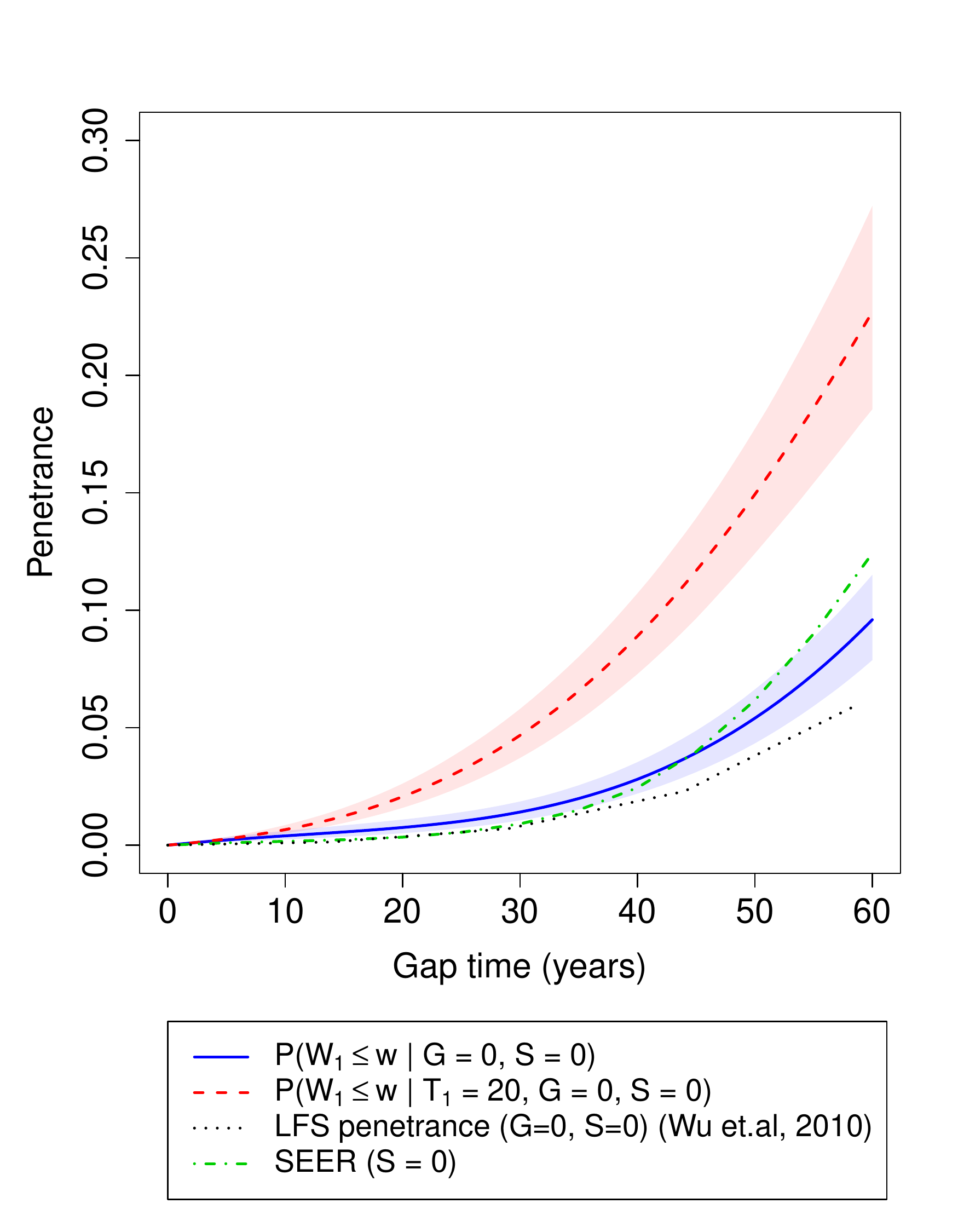}
		\label{fig:c}
	\end{subfigure} 
	\begin{subfigure}[]
		\centering
		\includegraphics[width=0.44\linewidth]{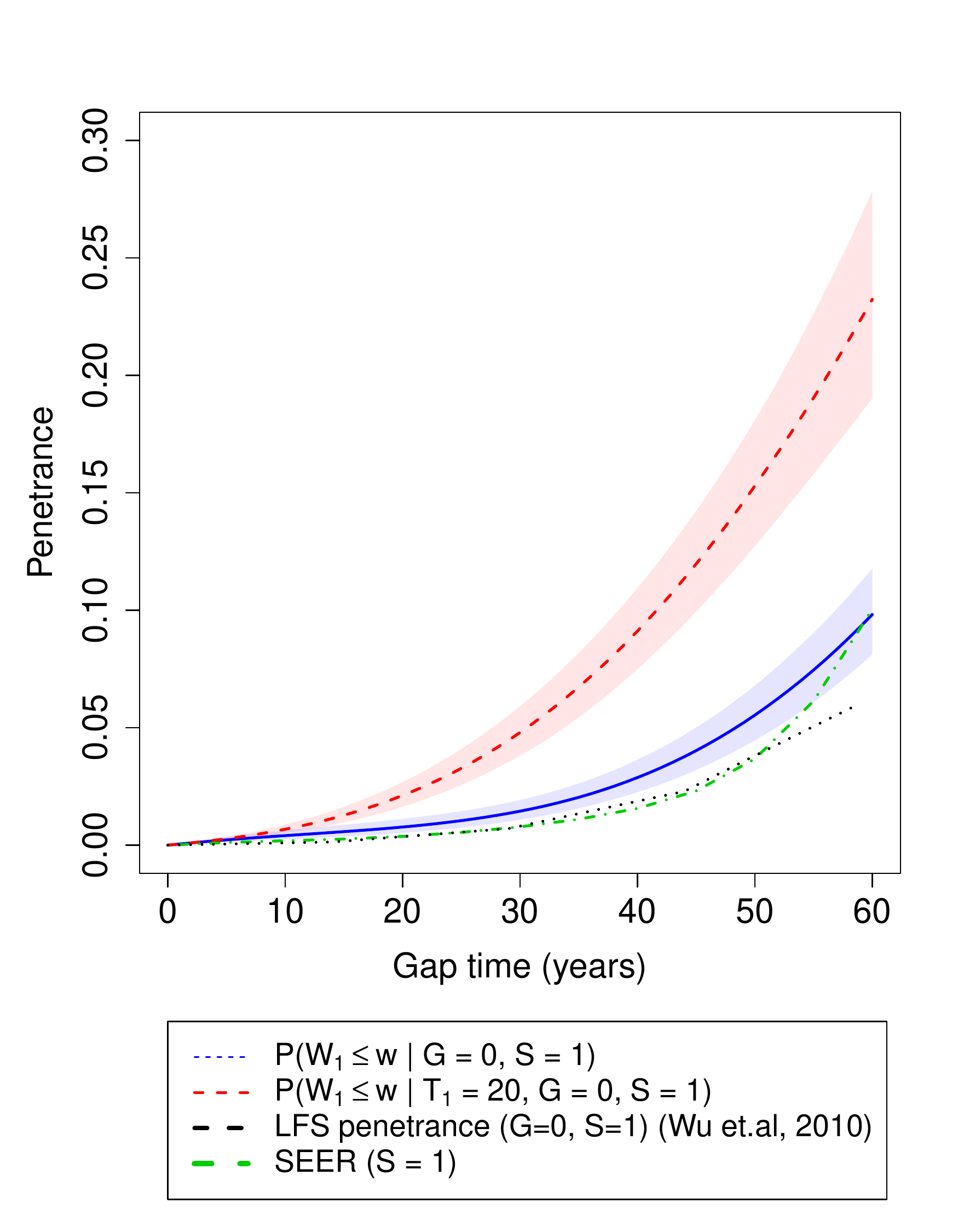}
		\label{fig:d}
	\end{subfigure}
	\caption{Age-at-onset penetrances of SPC and MPC for
            (a) female mutation carriers, (b) male mutation carriers,
            (c) female mutation non-carriers and (d) male mutation
            non-carriers. The shaded area is the 95\% credible
            bands. Note the y-axis scales between carriers and
            non-carriers are different. Notations: $G=1$, mutation
            carriers; $S=1$, male; $T_1=20$, the first primary cancer
            diagnosed at age 20. ``LFS penetrance'' denotes an
            estimate for 
                $\Pr\left\{\mbox{First primary cancer diagnosis} \mid \mbox{TP53
                    mutation status} \right\}$ that was previously published using a subset of our LFS dataset without
                considering the onset of multiple primary cancers.} \label{fg::penetranceEstimates}
\end{figure}

Figure {\ref{fg::penetranceEstimates}} compares penetrance estimates at different ages for females and males, respectively stratified by genotype. As expected, \textit{TP53} mutation has a clear effect on the increase of cancer risk, especially when the individual has a recent history of cancer. For an individual without a \textit{TP53} mutation, a history of cancer also has a positive effect on increasing the risk of developing subsequent cancer. 

\citet{WuStrongShete2010} estimated \textit{TP53} penetrance for the first primary cancer only from six pediatric sarcoma families, a subset of our LFS dataset. Figure {\ref{fg::penetranceEstimates}} shows that, for mutation carriers, this age-at-onset \textit{TP53} penetrance estimate aligns with those for SPC in our model but is slightly increased at an older age. Such consistency with a published analysis validates the performance of our model in real data. Another validation is when we compared our estimates for non-carriers to population estimates from the Surveillance, Epidemiology, and End (SEER) Results program \citep{lag1975seer}, they also align well (Figures \ref{fg::penetranceEstimates}c and d, more consistent in males than in females).

\subsection{Cancer risk prediction}
\begin{figure} 
\begin{center}
\centerline{
\includegraphics[width = 4in]{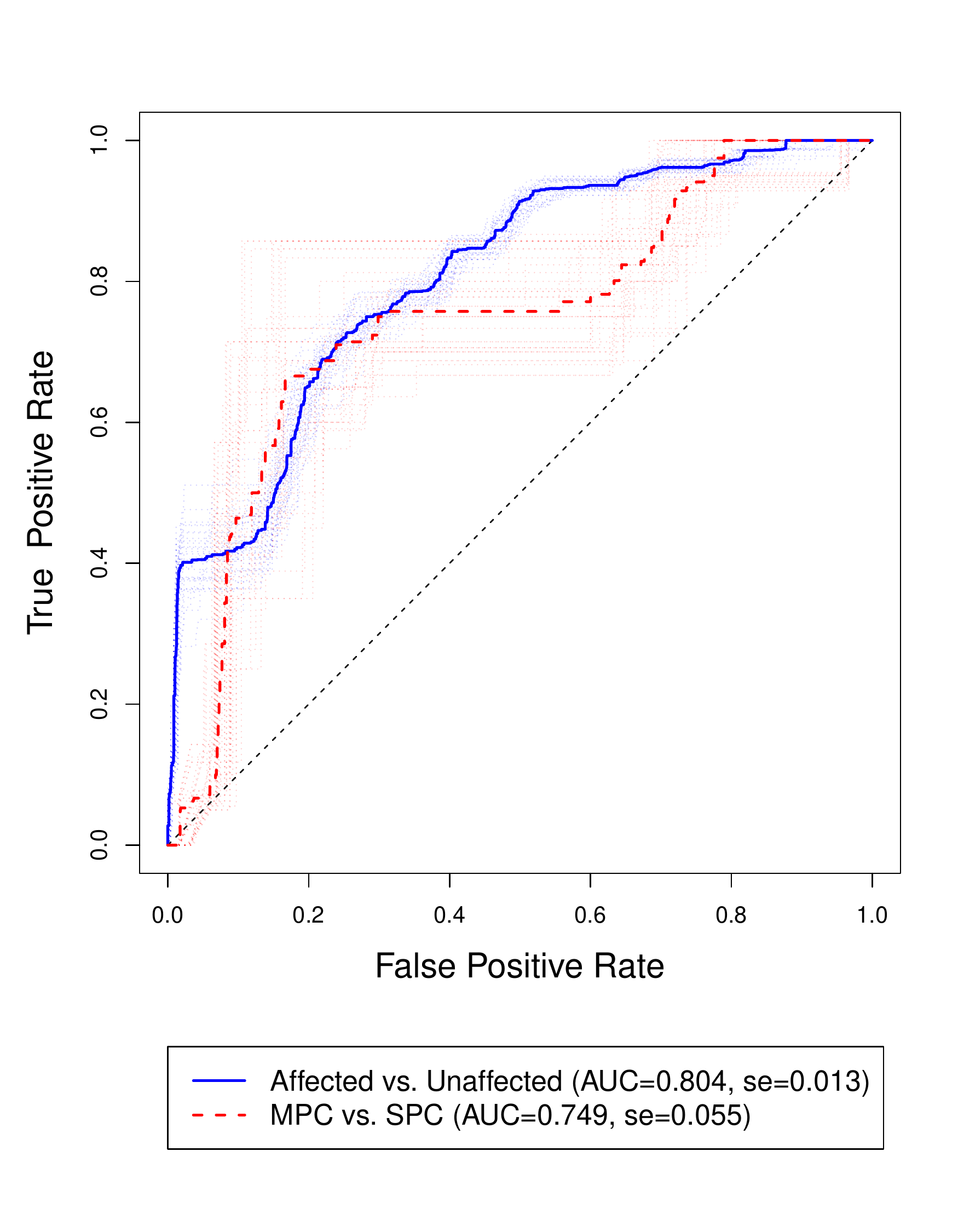}
}
\caption{ROC of the 5-year risk of developing the second primary
  cancer in the LFS dataset. The dotted lines denote the ROC curves
  for 25 random splits of the data, each undergone a 10-fold
  cross-validation. The solid lines denote the median ROC
  curves. \textit{Affected vs. Unaffected}, prediction of
    developing a cancer given that the individual has no history of
    cancer; \textit{MPC vs. SPC}, prediction of developing the next
    cancer given that the individual has had one primary
    cancer (SPC). Sample size: n(Affected)=123, n(Unaffected)=643, n(MPC)=21, n(SPC)=33. Abbreviation: se, standard error. }
\label{fg::cancerRiskPrediction}
\end{center}
\end{figure}

We assessed the ability of our model to predict cancer risk prediction using 10-fold cross-validation. We randomly split the 189 families into 10 portions and repeatedly fit our model to the 9 of the portions of all the families to estimate the penetrance, based on which we made predictions using the remaining 1 portion of the data. The individuals used for prediction are those who have known genotype information. We removed the probands because they were not randomly selected for genotype testing. We rolled back five years from the age of diagnosis of cancer or the censoring age. Based on the rolled-back time, we then calculated a 5-year cumulative cancer risk. We made two types of risk predictions that are of clinical interest. In the first scenario, we predicted the 5-year risk of developing a cancer given that the individual has no history of cancer (affected versus unaffected). In the second scenario, we predicted the risk of developing the next cancer when the individual had developed cancer previously (SPC versus MPC). We combined these results with those from the 10-fold cross-validation together and evaluated them using the receiver operating characteristic (ROC) curves. To assess the variation in prediction caused by data partitioning, we performed random splits for cross-validation 25 times. Figure {\ref{fg::cancerRiskPrediction}} shows the risk prediction results from each random split. The median area under the ROC curve (AUC) is 0.804 for predicting the status of being affected by cancer versus the status of not being affected by cancer, given that the individual has no history of cancer. The median AUC is 0.749 for predicting the status of the next cancer when the subject has had one primary cancer. The validation showed that the model performance is robust to random splits in cross-validation.

\section{Discussions} \label{s::discussion}

To our knowledge, this is the first attempt to estimate MPC-specific penetrance for \textit{TP53} germline mutation to include family members with unknown genotype information, which will, in turn, substantially improve the sample size and power of a study. In our LFS study, the increases in the number of cancer patients used in the analysis are 33\% (from 27 to 40) for MPCs, and 47\% (from 274 to 518) for the control group of SPCs. We developed a novel NHPP and incorporated it with a familywise likelihood so that it can model MPC events in the context of a family, while properly accounting for age effects and time-varying cancer status. We applied a Bayesian framework to estimate the unknown parameters in the model. We also adjusted for ascertainment bias in the likelihood calculation so that our penetrance estimates can be compared to those generated from the general population. Our new method provides a flexible framework for the penetrance estimation of MPC data, and shows reasonable predictive performance of cancer risk. As the number of patients with MPC continues to rise in the general population, our method will be useful to predict subsequent cancers and to assist in clinical management of the disease.  

Some possible extensions remain. First, we restricted our analysis up to the second primary cancer because of the limited power in the LFS dataset. This makes our penetrance estimation unsuitable for individuals with a history of more than two cancers. It is straightforward to extend our model to account for three or more cancers if we have such cases for each subpopulation. 

Second, the occurrence of primary cancers may depend on other factors such as cancer treatment. For example, radiotherapy can damage normal cells in tumor-adjacent areas and is associated with excessive incidence of secondary solid cancers \citep{inskip2007new}. Our model can include additional covariates to adjust for such dependency between successive events. However, the availablity of reliable data on radiotherapy is scarce and we have shown here that the current model can have reasonable predictive performance even without incorporating treatment factors.  

Third, because the correlation between the first two gap times in the real data is very small, the recurrent event model we used in this study does not explicitly consider such an association. For future datasets that exhibit a stronger level of correlation between the gap times, we would expect the predictive performance for the second or subsequent primary cancers to be improved by properly utilizing such correlation information (i.e., through Bayesian parametric copula models for sequential gap time analyses \citep{meyer2015bayesian}). 

Finally, in MPC studies, there usually exist multiple types of cancer. In our LFS study, even though our genetic background is simple, \textit{TP53} germline mutations, the presentation of cancer outcomes is diverse. LFS is characterized by occurrence of many different cancer types, such as sarcoma, breast cancer and lung cancer. Patients with MPC are thus subject to the competing risk of multiple types of cancer. In our current model, we pool together all cancer types and do not address the onset of second primary cancer at any specific site. As we collect more datasets on LFS from multiple clinics to increase our sample size, future work will include extending our methodology to provide MPC-specific and cancer-specific penetrance estimation.  

Finally, we provide an illustration of our method in Appendix F (Supplementary Materials). The associated example dataset and results, and all of the source code, are available at \url{http://github.com/wwylab/MPC}.

\section*{Acknowledgements}

Jialu Li is supported in part by the Cancer Prevention Research Institute of Texas through grant number RP130090. Jasmina Bojadzieva and Louise C. Strong are supported in part by the U.S. National Institutes of Health through grant P01CA34936. Wenyi Wang is supported in part by the Cancer Prevention Research Institute of Texas through grant number RP130090, and by the U.S. National Cancer Institute through grant numbers 1R01CA174206, 1R01 CA183793 and P30 CA016672. {\it Conflict of Interest}: None declared.

\bibliographystyle{biorefs}
\bibliography{references}

\newpage

\newcolumntype{C}{>{\centering\arraybackslash\columncolor{gray!10}}X}
\begin{table}
\caption{Number of primary cancers in the LFS dataset} \label{tb::numprimary}
\centering


\begin{tabularx}{\columnwidth}{c@{\extracolsep{\fill}} CCC C}


\hline
\rowcolor{white}  
Number of primary cancers & Gender  & Wildtype & Mutation & Unknown \\ 
\hline
\rowcolor{white}  
0  & Male  & 295 & 9 & 1276 \\
\rowcolor{white}  
   & Female  & 341 & 8 & 1207 \\

\hline
\rowcolor{white}  
1 & Male  & 105 & 25 & 139 \\
\rowcolor{white}  
   & Female  & 121 & 23 & 105 \\
   
\rowcolor{white} 
2 & Male  & 3 & 9 & 8 \\
\rowcolor{white}  
   & Female  & 3 & 12 & 5 \\

\rowcolor{white}    
3 & Male  & 0 & 3 & 0 \\
\rowcolor{white}  
   & Female  & 0 & 2 & 2 \\

\rowcolor{white}    
4 & Male  & 0 & 2 & 0 \\
\rowcolor{white}  
   & Female  & 0 & 1 & 0 \\

\rowcolor{white}    
5 & Male  & 0 & 0 & 0 \\
\rowcolor{white}  
   & Female  & 0 & 1 & 0 \\

\rowcolor{white}    
7 & Male  & 0 & 0 & 0 \\
\rowcolor{white}  
   & Female  & 0 & 1 & 0 \\

\hline

\rowcolor{white}    
\multicolumn{2}{c}{Total number of individuals} & 868 & 96 & 2742\\
\rowcolor{white}     
\multicolumn{2}{c}{Total number of cancer patients} & 232 & 79 & 259 \\
\rowcolor{white}    
\multicolumn{2}{c}{Total number of MPC patients}& 6 & 31 & 15 \\ 
\hline  

\end{tabularx}
\end{table}





%

\begin{table}
\caption{Summary of posterior estimates. sd, standard deviation.} \label{tb::posteriorEstimates}
\centering


\begin{tabular*}{\columnwidth}{c@{\extracolsep{\fill}} 
		c@{\extracolsep{\fill}} 
		c@{\extracolsep{\fill}} 
		c@{\extracolsep{\fill}}
		c@{\extracolsep{\fill}} 
		c@{\extracolsep{\fill}} 
		c@{\extracolsep{\fill}} 
		c@{\extracolsep{\fill}} 
		c@{\extracolsep{\fill}}
		c@{}}  \hline
	
	\hline
	\multirow{2}*{coefficient}
	& \multicolumn{4}{c}{Frailty} &&  \multicolumn{4}{c}{No Frailty} \\ \cline{2-5} \cline{7-10}
	& Median  & sd     & 2.5\%  & 97.5\% &&  Median    & sd       & 2.5\% & 97.5\% \\ 
	\hline
	$\beta_G$  				&  3.516 & 0.256 &  3.068 & 3.953 &&  3.288 & 0.223 &  2.871 & 3.687 \\
	$\beta_S$  				&  0.027 & 0.115 & -0.189 & 0.232 &&  0.027 & 0.118 & -0.187 & 0.241 \\
	$\beta_{G\times S}$ 	& -0.332 & 0.246 & -0.809 & 0.139 && -0.354 & 0.239 & -0.817 & 0.106 \\
	$\beta_{D(t)}$  		& -0.380 & 0.363 & -1.152 & 0.259 && -0.197 & 0.336 & -0.929 & 0.389 \\
	$\beta_{G\times D(t)}$  &  0.716 & 0.429 & -0.070 & 1.601 &&  0.700 & 0.402 & -0.033 & 1.548 \\
	$\phi$  		        &  5.883 & 1.695 &  3.427 & 9.855 && \multicolumn{4}{c}{Not Available}\\
	\hline
\end{tabular*}

\end{table}

\end{document}


\maketitle

\newpage

\appendix

\begin{center}
APPENDIX
\end{center}

\begin{center}
\section{Computation of IPCW Kendall's $\tau$}\label{appIPCW}
\end{center}

Letting $(X_1, Y_1)$ and $(X_2, Y_2)$ be two independent realizations of $(X, Y)$, the first and second gap times, and letting $\psi_{12} = I\{ (X_1- X_2)(Y_1- Y_2)>0 \}- I\{ (X_1- X_2)(Y_1- Y_2)<0 \}$ indicate the concordant/discordant status of the pair, the Kendall's $\tau$ \citep{gibbons1990rank} can be estimated from uncensored bivariate data $\{(X_i, Y_i), i=1, \dots, n \}$ by 
$$ {\binom{n}{2}}^{-1} \sum_{i<j} \psi_{ij} $$.
In the presence of censoring events $(V_X, V_Y)$, respectively related to the two gap times, the estimation of $\tau$ can only be based on orderable pairs. Let one observation be denoted as $(\tilde{X}, \tilde{Y}, \delta_X, \delta_Y )$, where $\tilde{X}= \min(X, V_X)$, $\tilde{Y}= \min(Y, V_Y)$, $\delta_X= I(X<V_X)$ and $\delta_Y= I(Y<V_Y)$. \citet{oakes1982concordance} showed that the pair $(i,j)$ is orderable if $\{\tilde{X}_{ij}< \tilde{V}_{X_{ij}}, \tilde{Y}_{ij}< \tilde{V}_{Y_{ij}} \}$, where $\tilde{X}_{ij} = \min(X_i, X_j)$, $\tilde{Y}_{ij} = \min(Y_i, Y_j)$, $\tilde{V}_{X_{ij}} = \min({V_X}_i, {V_X}_j)$, and $\tilde{V}_{Y_{ij}} = \min({V_Y}_i, {V_Y}_j)$. Letting $L_{ij}$ be the indicator of this event, and $\hat{p}_{ij}$ be an estimator of the probability of being orderable $p_{ij}= \Pr(V_X> \tilde{X}_{ij}; V_Y> \tilde{Y}_{ij} | \tilde{X}_{ij}, \tilde{Y}_{ij}) $, \citet{lakhal2010inverse} proposed the weighted estimate as 
$$\hat{\tau}_m = \left(\sum_{i<j} \frac{L_{ij}}{\hat{p}_{ij}} \right)^{-1} \sum_{i<j} \frac{L_{ij} {\psi}_{ij}} {\hat{p}_{ij}}  $$
To identify orderable pairs and estimate the corresponding $p_{ij}$,
\citet{lakhal2010inverse} showed that $L_{ij}$ can be reduced to that
$X_i$ and $X_j$ being uncensored, $\tilde{Y}_{ij}$ being observed, and that $\{ {V_X}_i> X_i+ \tilde{Y}_{ij}; {V_X}_j> X_j+ \tilde{Y}_{ij} \}$. The conditional probability of a pair being orderable is then
\begin{eqnarray*}  \label{path}
	p_{ij} &=& \Pr\{ {V_X}_i> X_i+ \tilde{Y}_{ij}; {V_X}_j> X_j+ \tilde{Y}_{ij} | X_i, X_j, \tilde{Y}_{ij} \}
	\\
	&=& G(X_i+ \tilde{Y}_{ij}) \times G(X_j+ \tilde{Y}_{ij})
\end{eqnarray*}            
The probability is estimated by 
$$\hat{p}_{ij} = \hat{G}(X_i+ \tilde{Y}_{ij}) \times \hat{G}(X_j+ \tilde{Y}_{ij})$$
where $\hat{G}(.)$ is the Kaplan-Meier estimator of $G(.)$ based on
$\{ (\tilde{X_k} + \tilde{Y_k}, 1-\delta_{Y_k}), k=1, \cdots, n
\}$. The standard error of the Kendall's $\tau$ is estimated by the
jackknife technique. 

\section{An example of using the peeling algorithm to calculate the familywise likelihood}\label{appPeeling}

\begin{figure} 
\centerline{\includegraphics[width = 3in]{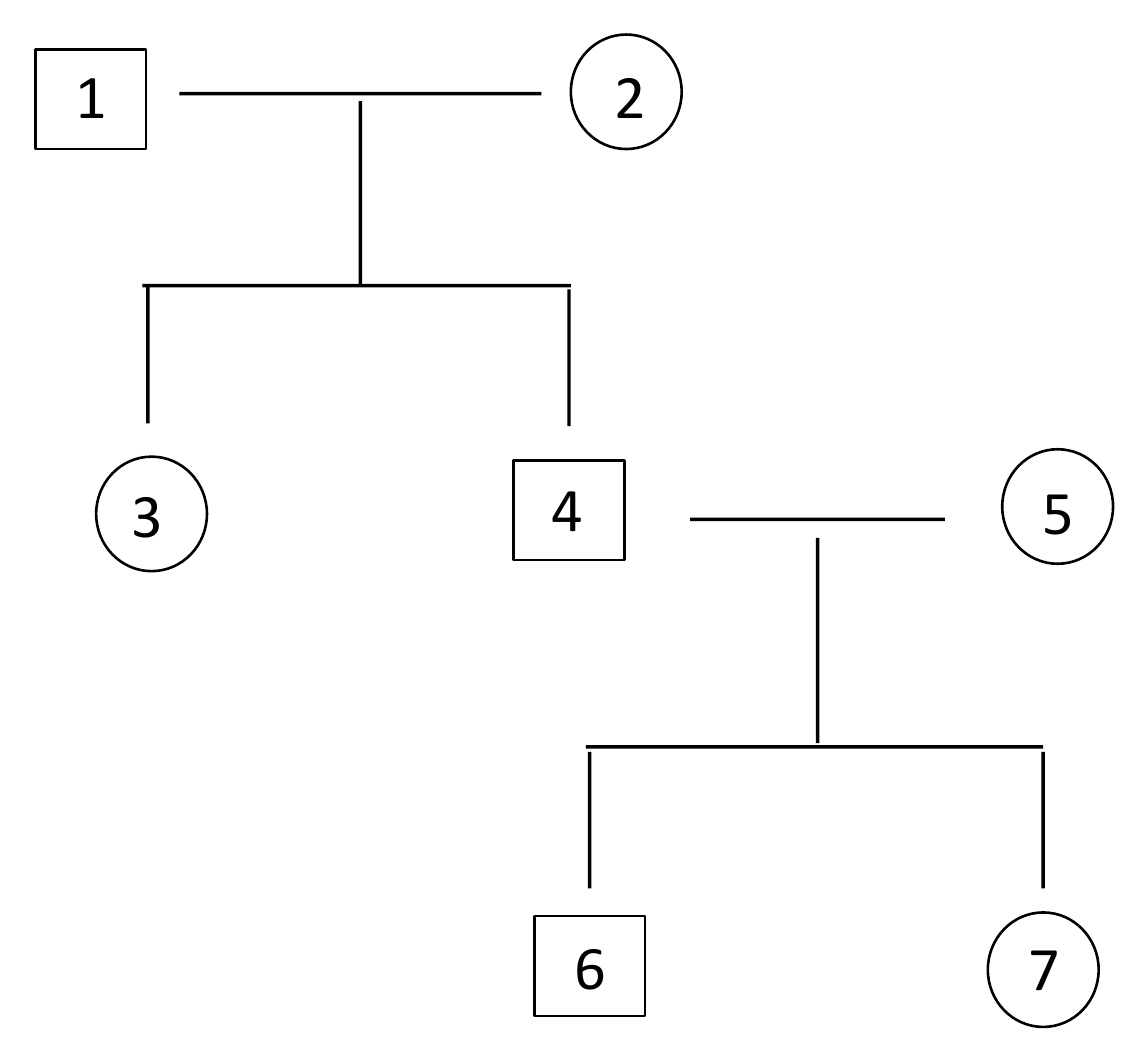}}
\caption{Supplementary Figure 1. A hypothetical pedigree to illustrate the likelihood
  calculation using the Elston-Stewart algorithm. The family consists
  of three generations. The circle indicates the female member while
  the square indicates the male. The horizontal lines indicate
  marriage and vertical lines indicate the next generation. In this example, the genotype is
  assumed unknown for every members except the 1st and 4th individuals.}\label{fg::peelingPedigree} \vspace*{-3pt}
\end{figure}

Supplementary Figure {\ref{fg::peelingPedigree}} shows an example of a hypothetical
family with 3 generations. Without loss of generality, we assume that
$\bg_{obs}^T = (g_1, g_4)$ and let $\bg^T_{mis}= (g_2, g_3, g_5, g_6,
g_7)$ and $\bH^T= (h_1, \cdots, h_7)$ denote vectors of the unknown
genotypes and the cancer history of the family, respectively. The
peeling algorithm peels through the family by considering individuals 1, 2, 3 as anterior and
individuals 5, 6, 7 as posterior of individual 4. We can then compute the family-wise likelihood $\Pr(\bh | \bg_{obs})$ as follows:
\begin{align*}
\Pr(\bh & | \bg_{obs}) \\
& = \Pr(h_4 | \bg_{obs}) \times \Pr(h_1, h_2, h_3 | \bg_{obs}) \times \Pr(h_5, h_6, h_7 | \bg_{obs}) \\
& = \Pr(h_4 |g_4) \times \Pr(h_1 | g_1) \cdot \Pr(h_2, h_3 | g_1, g_4) \times 
    \Pr(h_5, h_6, h_7  | g_1, g_4) \\
& = \Pr(h_4 |g_4) \times \Pr(h_1 | g_1) \cdot \left[\sum_{g_2}\Pr(h_2|g_2) \Pr(h_3 | g_1, g_2, g_4) \Pr(g_2|g_1, g_4) \right] \\
& \qquad \times \left[\sum_{g_5} \Pr(h_5|g_5) \Pr(h_6, h_7| g_1, g_4, g_5) \Pr(g_5 | g_1, g_4)\right] \\
& = \Pr(h_4 |g_4) \times \Pr(h_1 | g_1) \cdot \left[\sum_{g_2}\Pr(h_2|g_2) \Pr(g_2|g_4)
\left\{\sum_{g_3} \Pr(h_3 |g_3) \Pr(g_3|g_1, g_2, g_4) \right\} \right] \\
& \qquad \times \left[\sum_{g_5} \Pr(h_5|g_5) \Pr(g_5) 
\left\{ \sum_{g_6} \Pr(h_6|g_6) \Pr(h_7|g_4, g_5) \Pr(g_6|g_4, g_5) \right\}\right] \\
& = \Pr(h_4 |g_4) \times \Pr(h_1 | g_1) \cdot \left[\sum_{g_2}\Pr(h_2|g_2) \Pr(g_2|g_4)
\left\{\sum_{g_3} \Pr(h_3 |g_3) \Pr(g_3|g_1, g_2, g_4) \right\} \right] \\
& \qquad \times \left[\sum_{g_5} \Pr(h_5|g_5) \Pr(g_5) 
\left\{ \sum_{g_6} \Pr(h_6|g_6) \Pr(g_6|g_4, g_5)
\left(\sum_{g_7} \Pr(h_7|g_7) \Pr(g_7|g_4, g_5) \right) \right\}\right].
\end{align*}
All probabilities in the last equation are straightforward to compute when the mode of inheritance is known.


\section{Bayesian estimation procedure}\label{mcmcProcedure}
In this study, we used the MCMC algorithm to generate posterior
distributions for model parameter estimation. The algorithm integrates
the Metropolis-Hastings algorithm, which draws posterior samples by
comparing posterior densities from two adjacent iterations, with the
Gibbs sampling scheme, which allows for sampling multiple model
parameters within an iteration by utilizing the full conditional
likelihood. More details about the MCMC algorithm can be found in
\citet{hoff2009first, gelman2014bayesian}. Here, we show the Bayesian
inference in the frailty model. The inference of the final model we used
for the LFS study can be made by simply removing the part for the frailty estimation.

Supplementary Figure \ref{fg::graphFrailtyModel} shows the frailty model represented by a directed graph that connects the observed data, model parameters and the hyper-parameter, and details about MCMC algorithm is summarized in the following: 

\begin{figure}
\centerline{
\includegraphics[width = 2.6in]{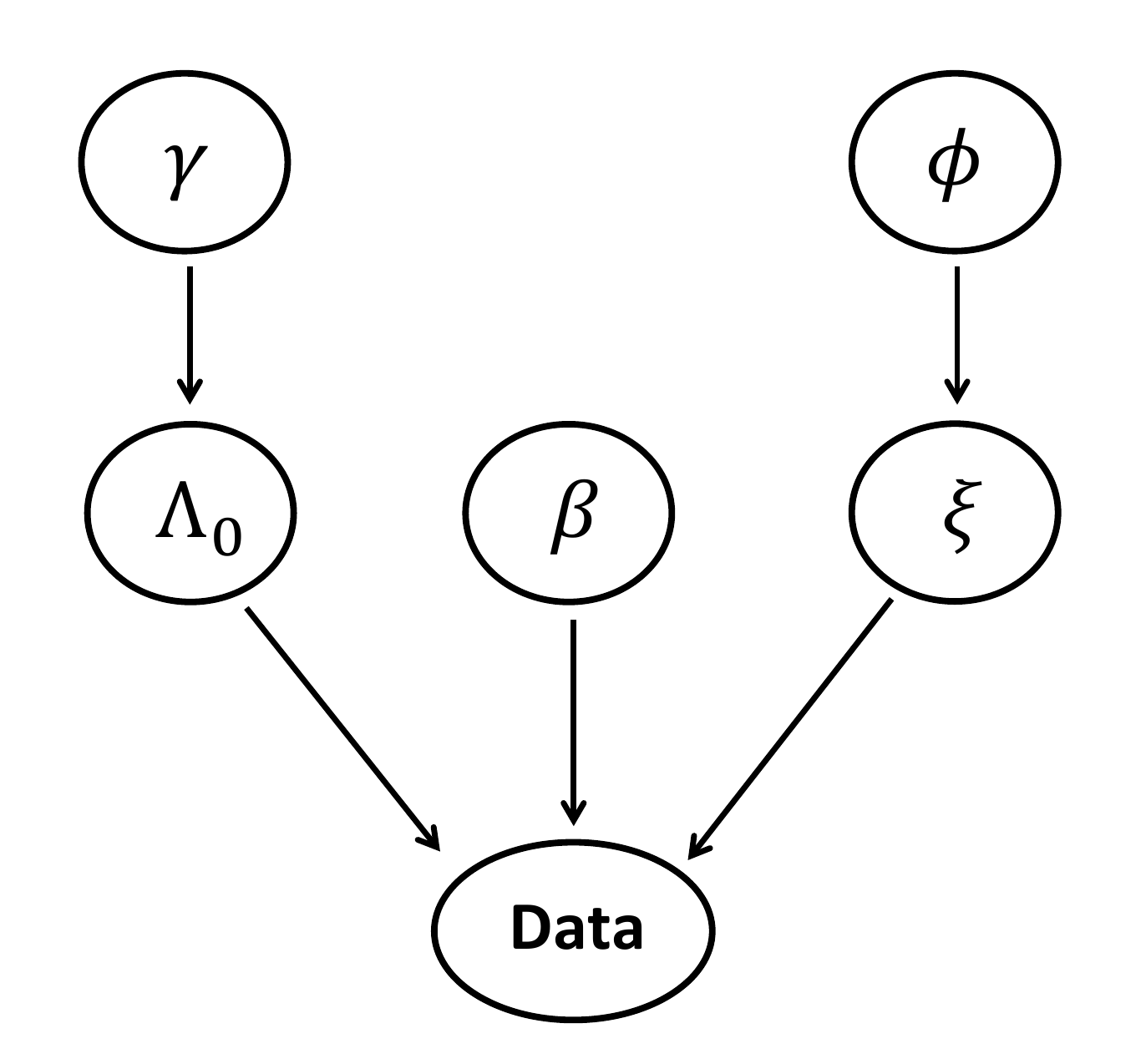}
}
\caption{Supplementary Figure 2. Graphical representation of the Bayesian frailty model. $\Lambda_0$ is the cumulative baseline rate function; $\phi$ is the hyper-parameter of frailty $\xi$. } \label{fg::graphFrailtyModel}
\end{figure}

\begin{itemize}
	\item {\bf Prior setting}\\
	$\beta \sim N(0, 100^2)$; 
	$\gamma$: flat prior;
	$\phi \sim Gamma(.01, .01)$ 
	
	\item {\bf Proposal setting} \\
	Given $\theta^{(t-1)}$, generate $\theta^* \sim q(\theta^{(t-1)})$ 
	 
	\item {\bf Iterative updating:} \\
	1) Compute proposal adjustment $adj= \frac{q(\theta^{(t-1)}| \theta^{*})}{q(\theta^{*}| \theta^{(t-1)})}$; \\  
	2) Let $\bh$ denote the cancer phenotype (or survival) data,
        and $p(\bh |\theta^*, others)$ denote the full conditional
        distribution of $\theta^*$, and compute the acceptance ratio 
	$$ r= min \bigg( \frac{p(\bh |\theta^*, others) p(\theta^*) } {p(\bh |\theta^{(t-1)}, others) p(\theta^{(t-1)})} * adj  , 1 \bigg)$$ 
	3) Take 
	\begin{equation*}
	\theta^{(t)}=
	\begin{cases}
	\theta^*,         & \text{with probability r}  \\
	\theta^{(t-1)},  & \text{with probability $1-r$}
	\end{cases}
	\end{equation*} 
	4) Sample $u \sim Uniform(0,1)$, and set $\theta^{(t)} = \theta^*$ if $u < r$ or $\theta^{(t)} = \theta^{(t-1)}$ otherwise.
	
\end{itemize}

Since we have parameters (e.g., $\gamma$, $\xi$ and $\phi$) that only take positive values, we employ a log-normal proposal. Suppose $\gamma^{(t-1)} \in (0, +\infty) \sim \log N(\mu, \sigma)$, and
$\log  \gamma^{(t-1)} \in (-\infty, +\infty) \sim N(\mu', \sigma')$. To propose a new sample, we generate $ \log  \gamma^*= \log  \gamma^{(t-1)} + \epsilon$ where $\epsilon \sim N(0,1)$, by which we can obtain $ \gamma^* = \exp(\log \gamma^*) \in (0, +\infty) $. To adjust the asymmetric proposal density, we calculate
$$adj= \frac{lnN (\gamma^{(t-1)} | ln\gamma^*) }{lnN (\gamma^* | ln\gamma^{(t-1)} )}= \frac{ \frac{1}{\gamma^{(t-1)} \sigma \sqrt{2\pi} } \exp[-\frac{(ln \gamma^{(t-1)} - ln \gamma^*)^2}{2 \sigma^2}] }
{\frac{1}{\gamma^* \sigma \sqrt{2\pi}} 
	\exp[-\frac{(ln \gamma^* - ln \gamma^{(t-1)})^2}{2 \sigma^2}] } = \frac{\gamma^*}{\gamma^{(t-1)}}$$
which is simply the ratio of the proposed samples.

The posterior density for $\phi$ was constructed as previously described \citep{clayton1991monte}. In brief, let $\phi \sim Gamma(\nu_a, \nu_b)$, or $f(\phi|\nu_a, \nu_b)= \frac{\nu_b^{\nu_a}}{\Gamma\{\nu_a\}} \phi^{\nu_a-1} \exp \left\{-\nu_b \phi \right\}$, where $\nu_a, \nu_b$ are the shape and rate of the Gamma distribution, respectively. The posterior density of $\phi$ is then

\begin{align*}
\Pr (\phi | \bxi) &\propto \Pr (\bxi | \phi) \Pr(\phi | \nu_a, \nu_b) \\ 
& = \prod_{i}^{I} \frac{\phi^{\phi} \xi_{i}^{(\phi-1)} \exp(-\phi \xi_i)}{\Gamma(\phi)} \frac{\nu_b^{\nu_a} \phi^{(\nu_a-1)} \exp(-\nu_b \phi)}{\Gamma(\nu_a)} \\
& = \frac{\phi^{I\phi + \nu_a -1} \exp(-\nu_b \phi) \exp \bigg(\bigg[(\phi-1) \log \prod_i^{I} \xi_i - \phi \sum_i^{I} \xi_i  \bigg] \bigg)} {\Gamma(\phi)^{I}}.
\end{align*} 

where $I$ denotes the number of families. \\

Finally, we implemented this MCMC algorithm in R as follows.
{\small{\verbatiminput{posterior.update.R}}}

\begin{figure}[!htbp]
\centerline{
\includegraphics[width = 5.5in]{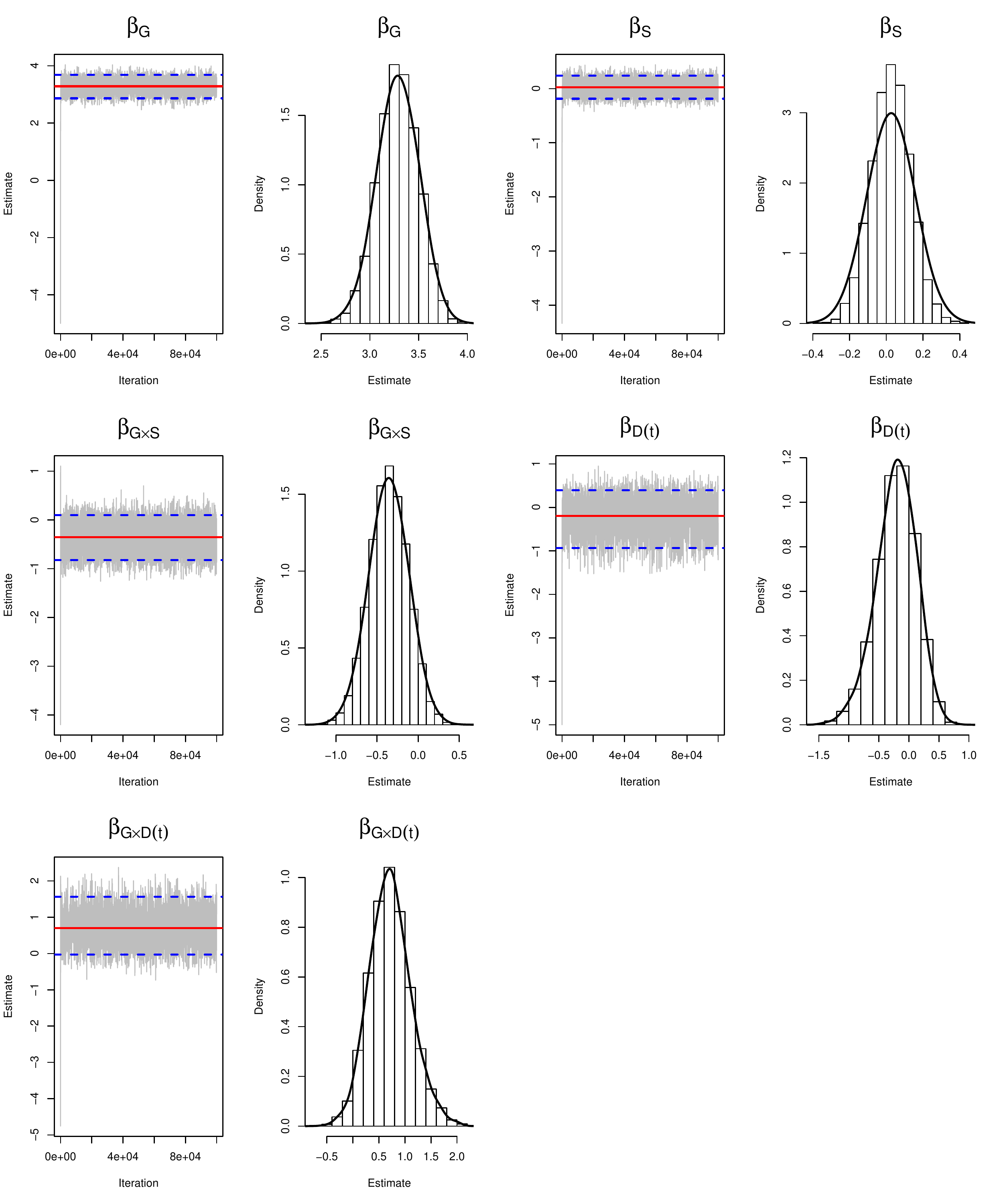}
}
\caption{Supplementary Figure 3. Trace plots and density distribution of posterior samples (after removing burn-in) from the proposed method. The red line indicates posterior median estimate. The density distribution is estimated based on the histogram. } \label{fg::mcmcDiagIndependentModel}
\end{figure}

\begin{figure}[!htbp]
\centerline{
\includegraphics[width = 5.5in]{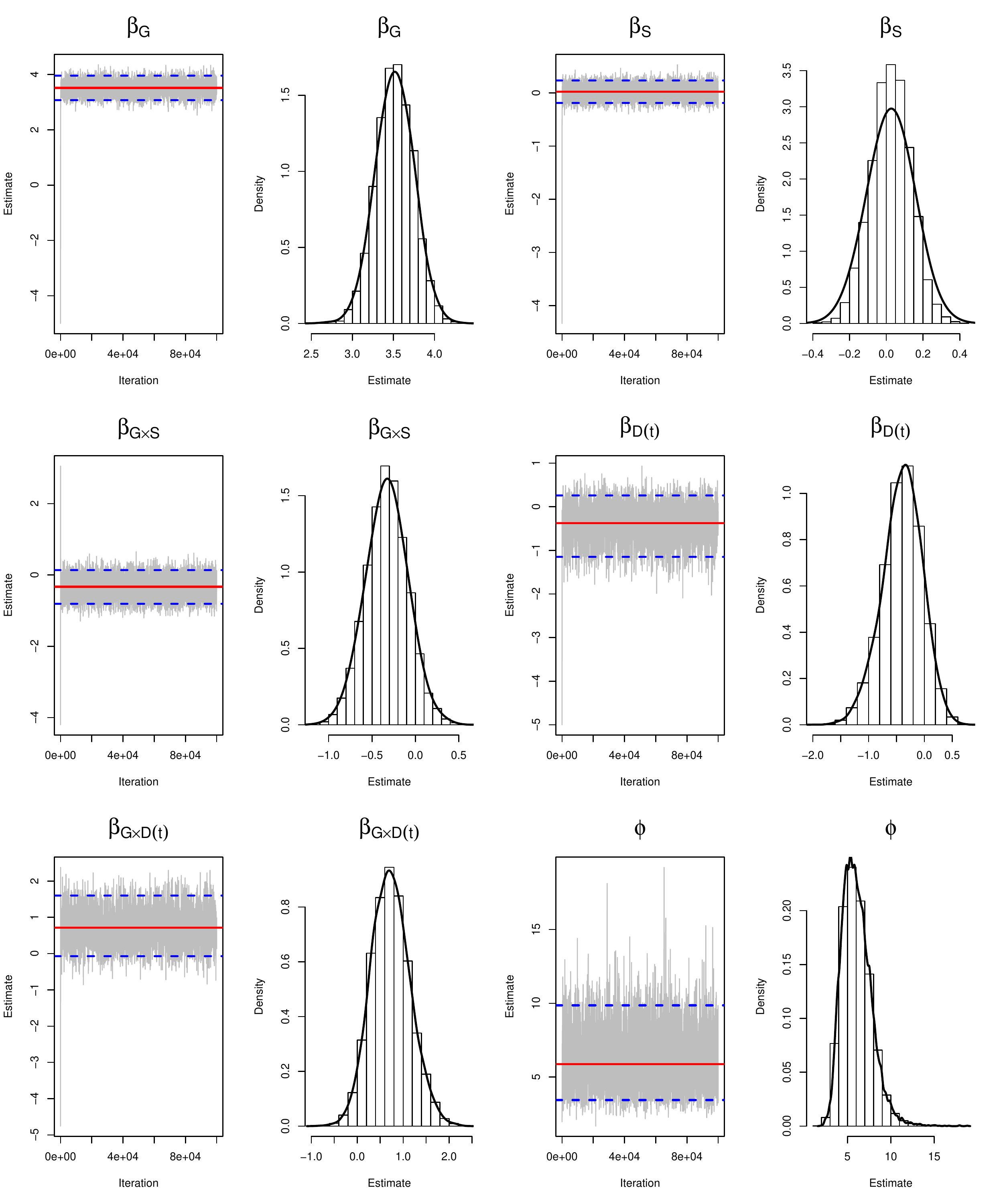}
}
\caption{Supplementary Figure 4. Trace plots and density distribution of posterior samples (after removing burn-in) from the frailty model. The red line indicates posterior median estimate. The density distribution is estimated based on the histogram.} \label{fg::mcmcDiagFrailtyModel}
\end{figure}

\begin{figure}[!htbp]
\centering
\subfigure[No Frailty]{
\includegraphics[width = 0.4\textwidth]{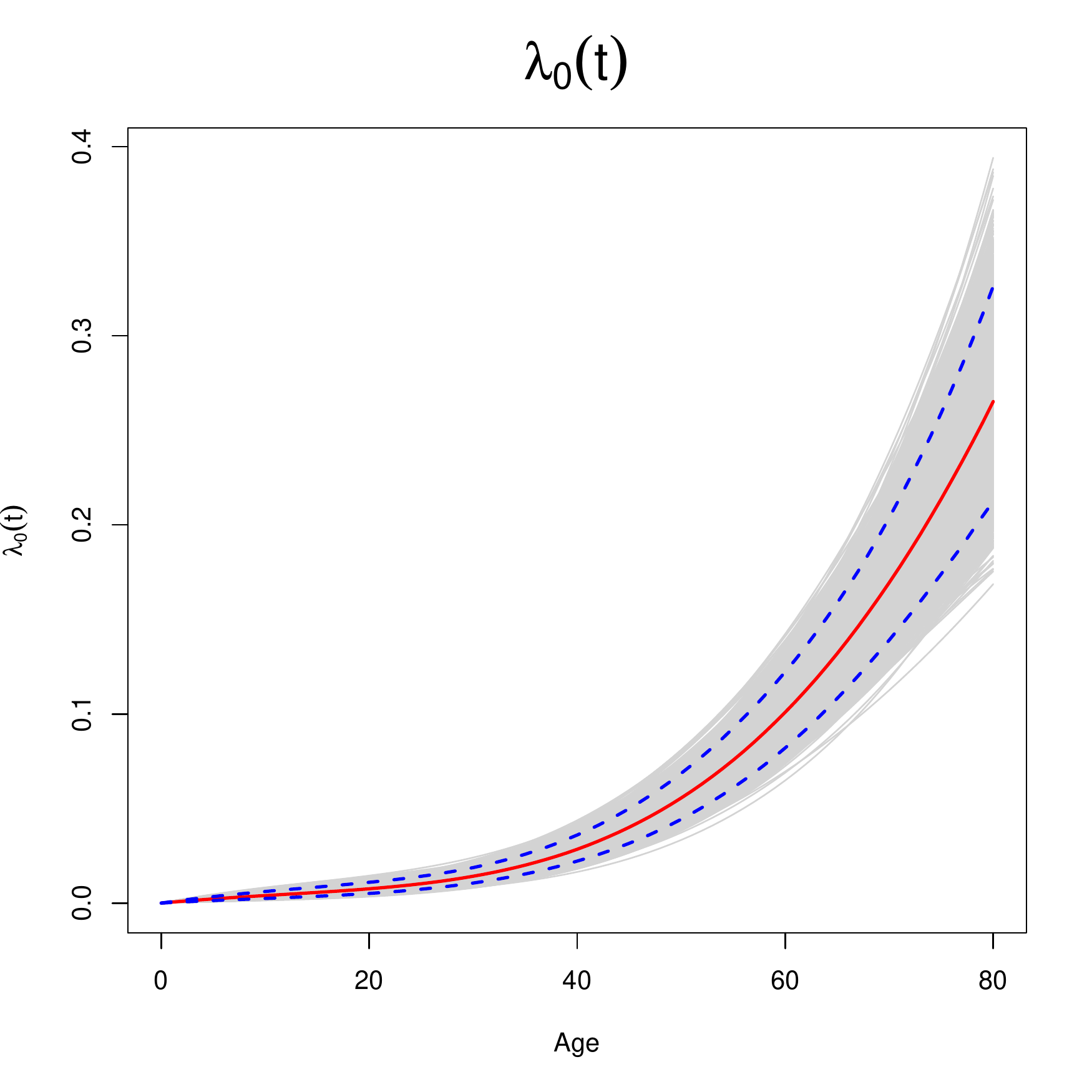}}
\subfigure[Frailty]{
\includegraphics[width = 0.4\textwidth]{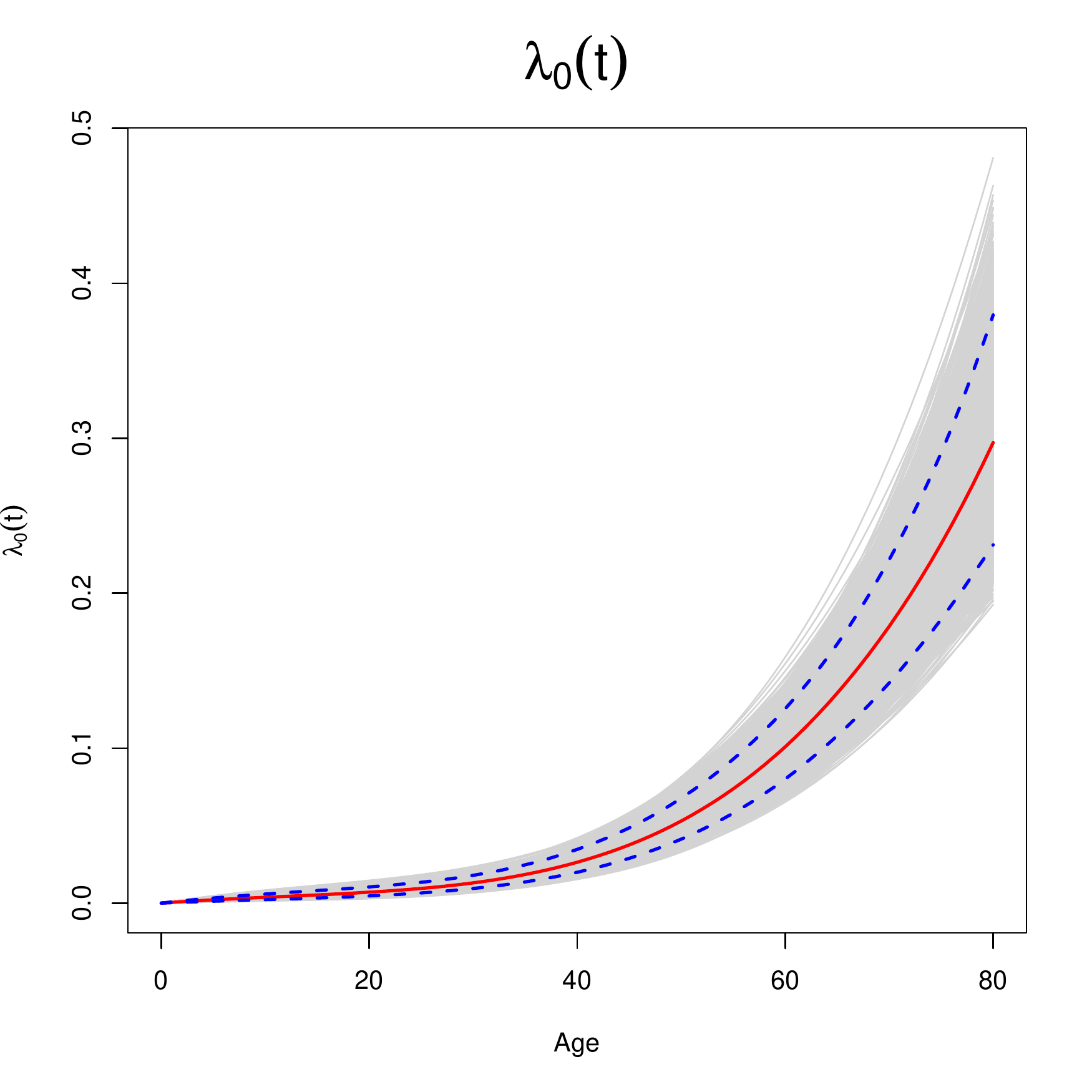}}
\caption{Supplementary Figure 5. Comparison of Baseline Estimates for frailty vs. no frailty models. } \label{base}
\end{figure}

To check the convergence of the algorithm, we applied the proposed
models both with and without frailty term to the real
data. Supplementary Figure
\ref{fg::mcmcDiagIndependentModel}, Supplementary Figure
\ref{fg::mcmcDiagFrailtyModel}, and Supplementary Figure \ref{base} show the results. Both models converges
well and the results are nearly identical.

\newpage

\section{Sensitivity prior analysis}\label{senseAnalysis}
\begin{figure}\newpage

\centerline{
\includegraphics[width = 3in]{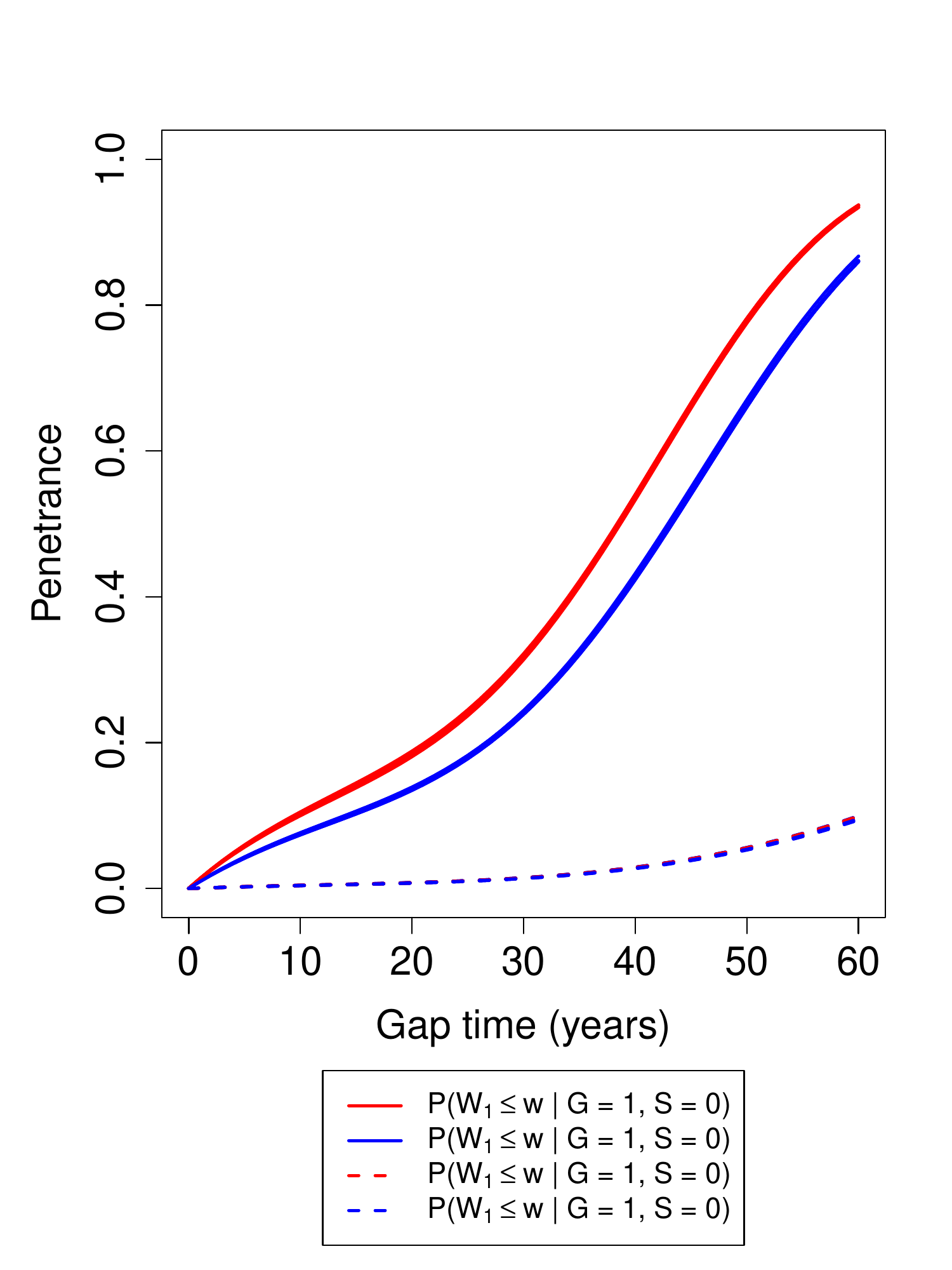}
\includegraphics[width = 3in]{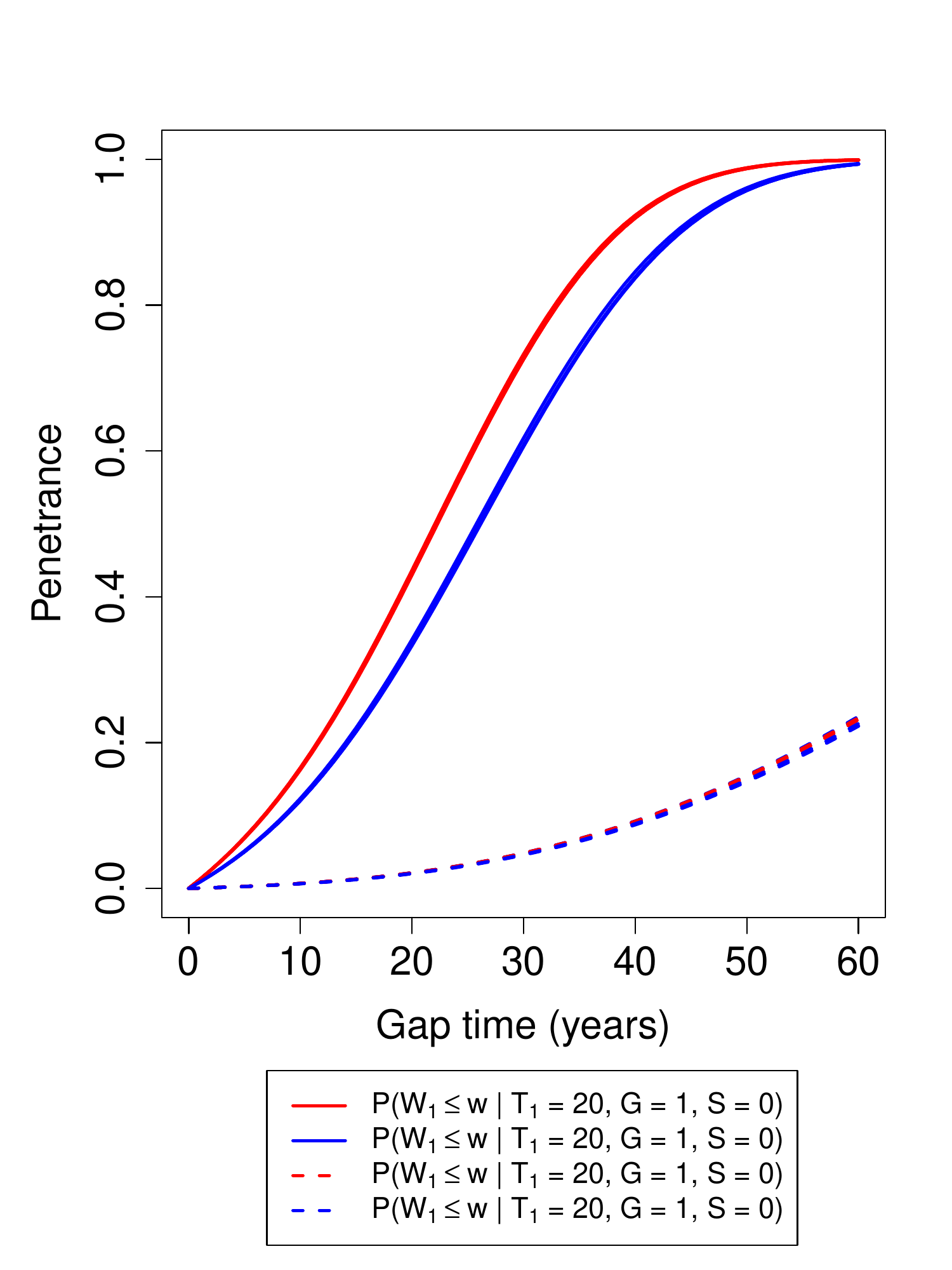}
}
\caption{Supplementary Figure 6. Penetrance estimates from sensitivity prior analysis for the first (left) or the second primary cancer (right). Penetrances estimated from the different combinations of prior settings are shown with the same color and line type for each subgroup.} \label{fg::sensitivityPriorAnalysis}
\end{figure}

We performed sensitivity analysis by comparing penetrance estimates
under different prior settings. We tested 6 combinations of priors for
$\beta$ and $\gamma$: three different priors for $\beta$, including
$Nomral(0, 100^2)$, $Normal(0, 10^2)$ and a flat prior, and three
different priors for $\gamma$ including $Gamma(0.1, 0.1)$ and a flat
prior. Supplementary Figure \ref{fg::sensitivityPriorAnalysis} shows their penetrance estimates for the first or the second primary cancers for each subgroup.        

\newpage
\section{Penetrance Estimates from the Frailty Model}\label{senseAnalysis}
Penetrance estimates from the frailty model and the model without
frailty are shown in Supplementary Figure \ref{fg::frailtyIndPenComparison}. There
is no obvious difference between the two sets of estimates.

\begin{figure}[!h]
\centerline{
\includegraphics[width = 3in]{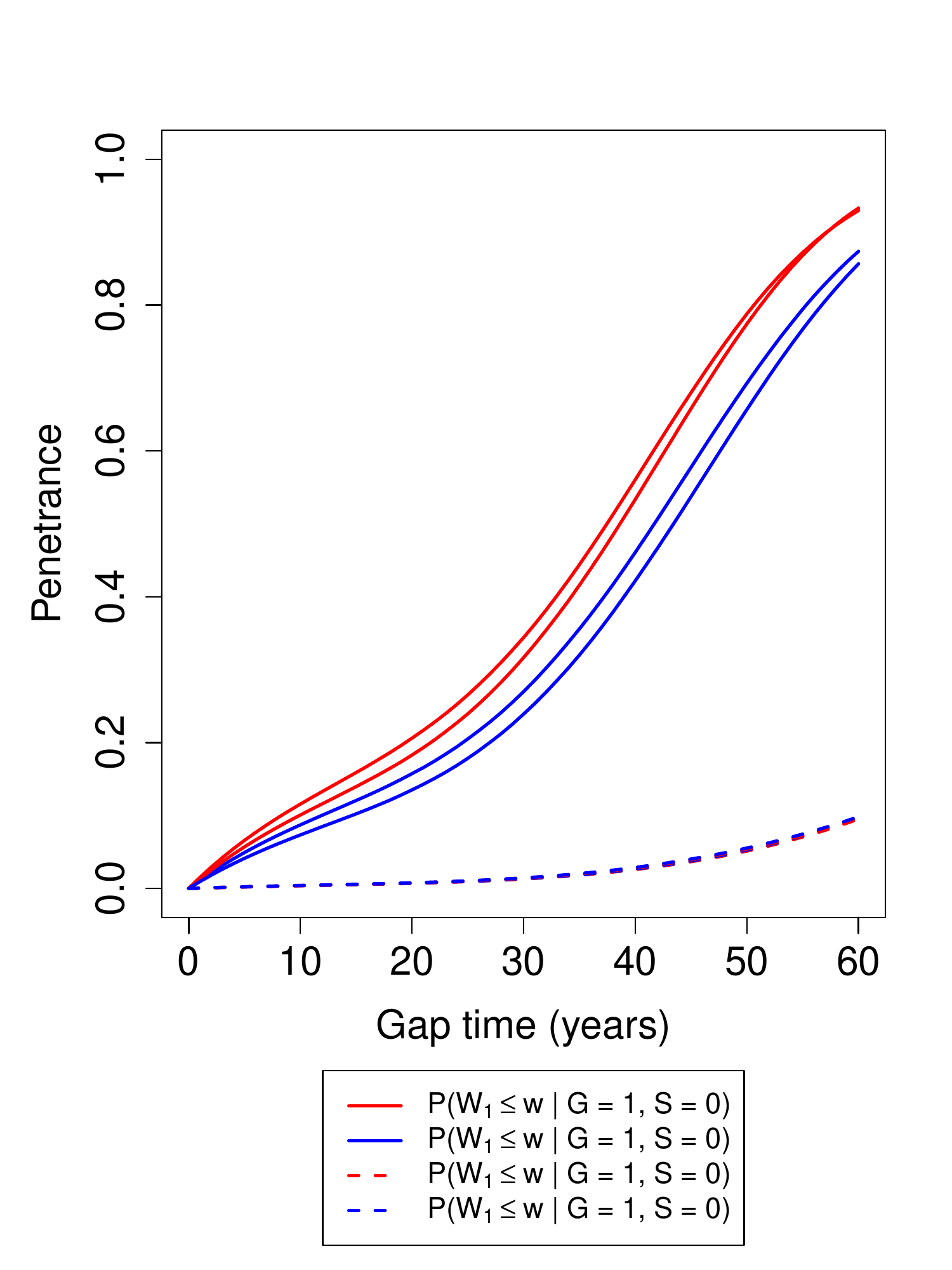}
\includegraphics[width = 3in]{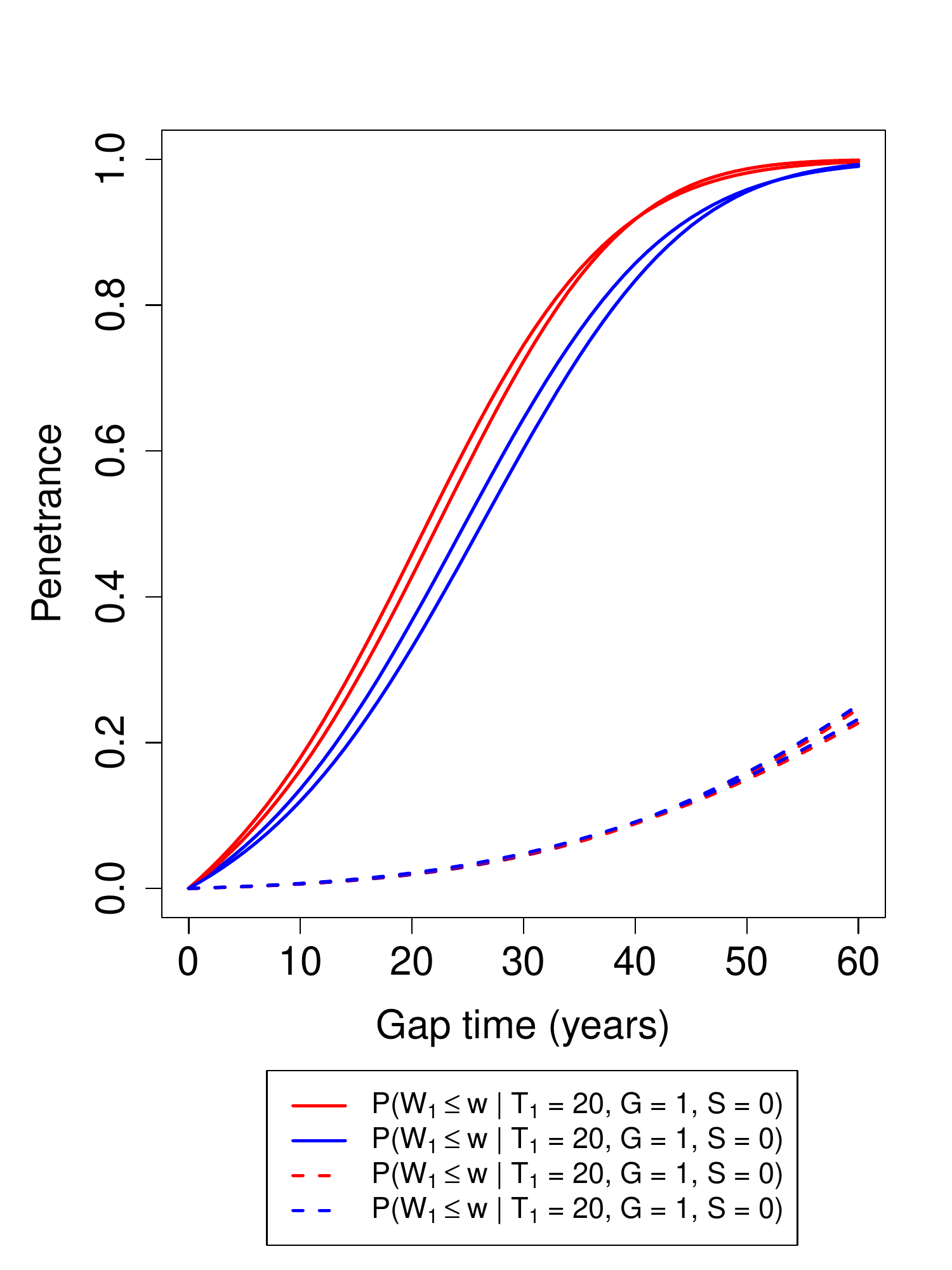}
}
\caption{Supplementary Figure 7. Comparison of penetrance estimates generated from frailty model and model without frailty.} \label{fg::frailtyIndPenComparison}
\end{figure}

\newpage
\section{Illustration of R-code}
We provide estimation results for a simulated dataset with 50
families. The data generation procedure is described in Section 4. As
shown in Supplementary Figure \ref{fg::illu}, our code successfully recovers the
true values of all parameters. The complete set of source code,
including the set that reproduces the results presented in this
section, is available at \url{http://github.com/wwylab/MPC}.
\begin{figure} [!htbp]
\begin{center}
\subfigure[$\beta_1$]{
\includegraphics[width = 0.4\textwidth]{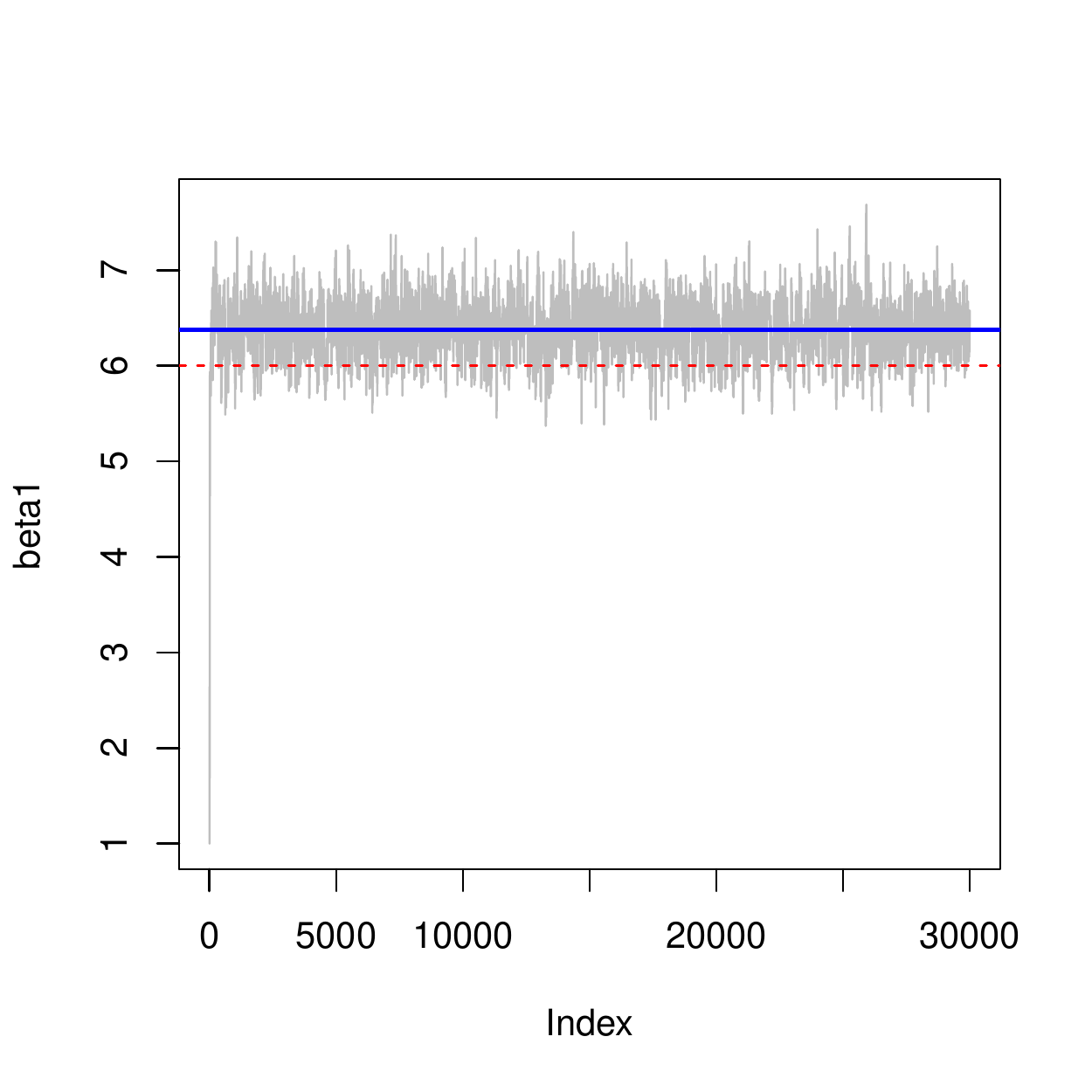}}
\subfigure[$\beta_2$]{
\includegraphics[width = 0.4\textwidth]{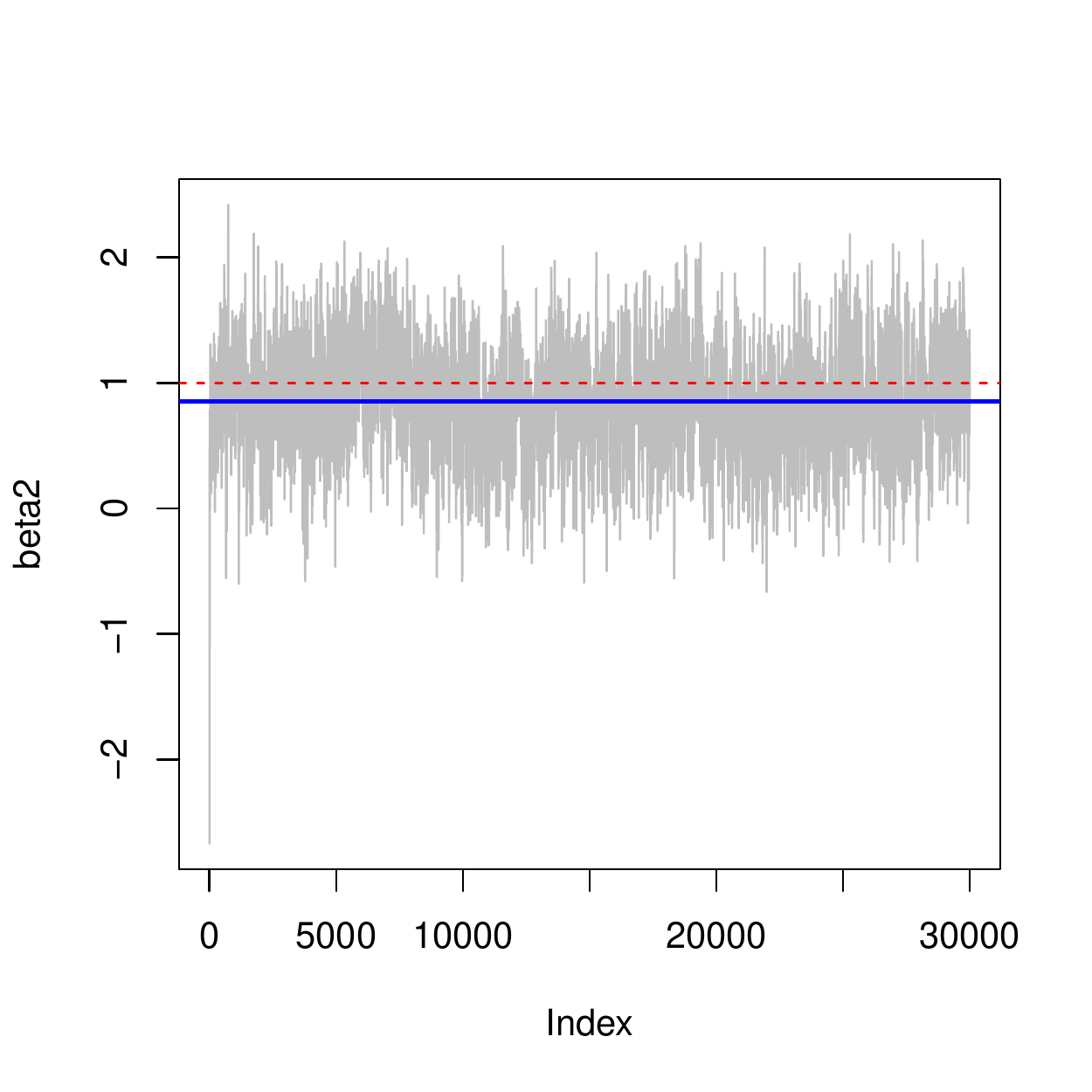}}
\subfigure[$\Lambda_0(t)$]{
\includegraphics[width = 0.4\textwidth]{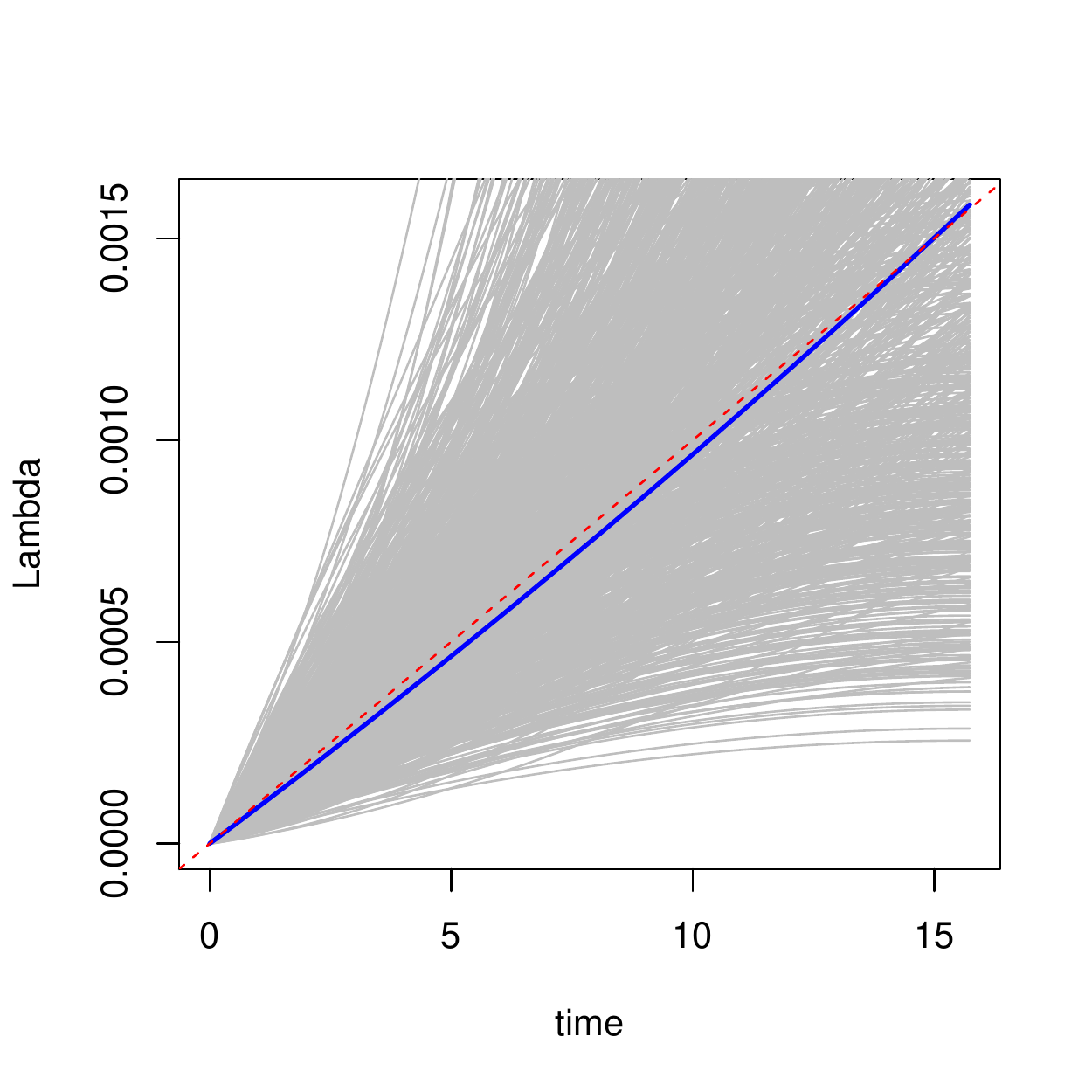}}
\subfigure[$\phi$]{
\includegraphics[width = 0.4\textwidth]{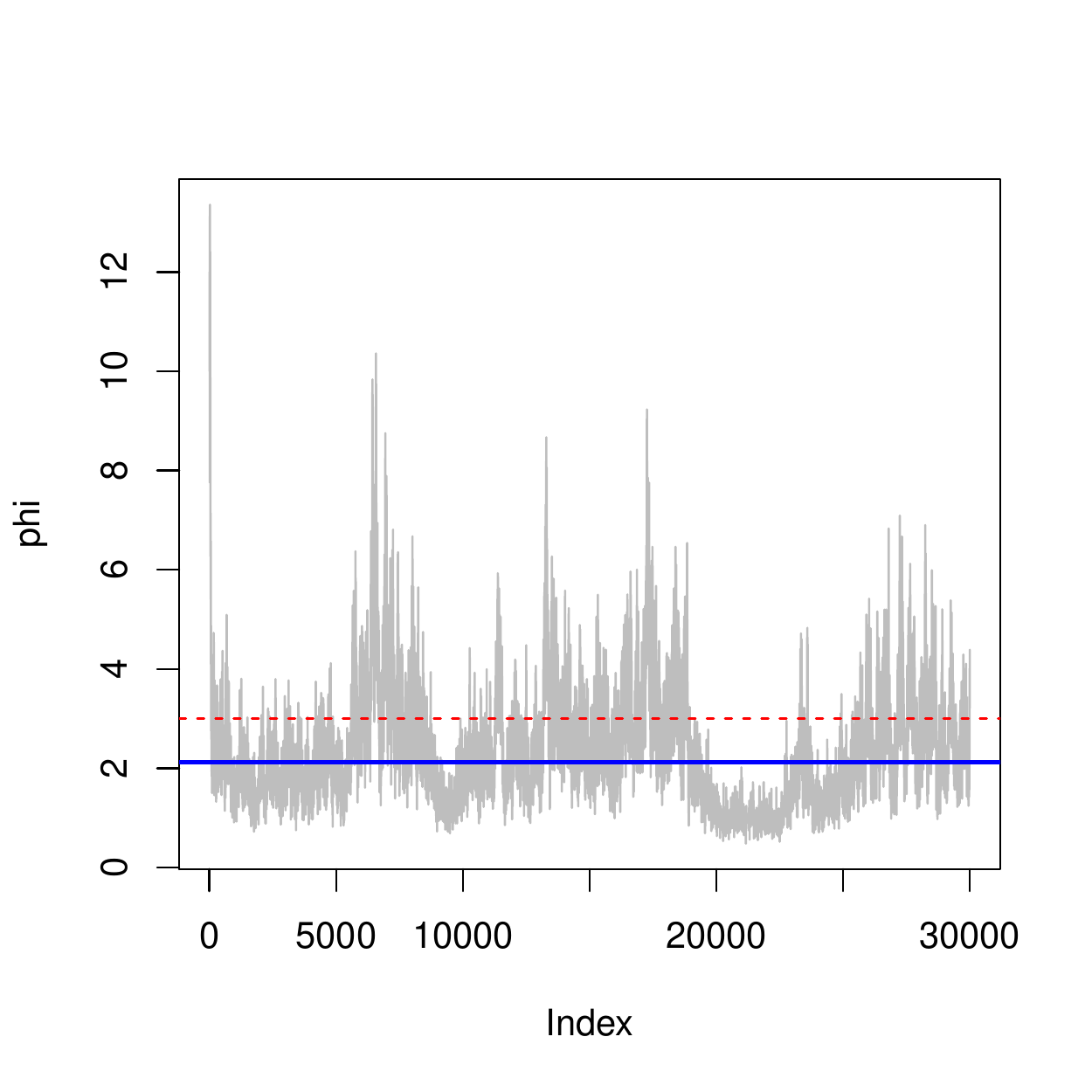}}
\end{center}
\caption{Supplementary Figure 8. Our code successfully recovers the true values of all parameters. Here (blue) solid lines represent posterior estimates and (red) dashed lines represent true values. } \label{fg::illu}
\end{figure}

\newpage

\section{Additional Supplementary Figures and Tables}
This section contains addition figures and tables referred to in the main manuscript of this article.  

\begin{table}[!htbp]
\caption{Supplementary Table 1. Summary of the LFS data referred in Section 2.1. "W/ carriers", family with at least one mutation carrier; "W/O carriers", family with no observed mutation carriers. } \label{tb::famsummary}
\centering
\begin{tabular*}{\columnwidth} {@{}l@{\extracolsep{\fill}} 
                                 c@{\extracolsep{\fill}} 
                                 c@{\extracolsep{\fill}} 
                                 r@{}} \hline
                      &  W/ carriers & W/O carriers & total \\ \hline
Number of families    & 17     & 172  &  189 \\
Number of individuals & 2,409   & 1,297 & 3,706 \\ \hline
Average family size   & 142 & 8 &  20 \\ \hline

\end{tabular*}
\end{table}

\begin{figure} [!htbp]
\centerline{
\includegraphics[width = 0.5\textwidth]{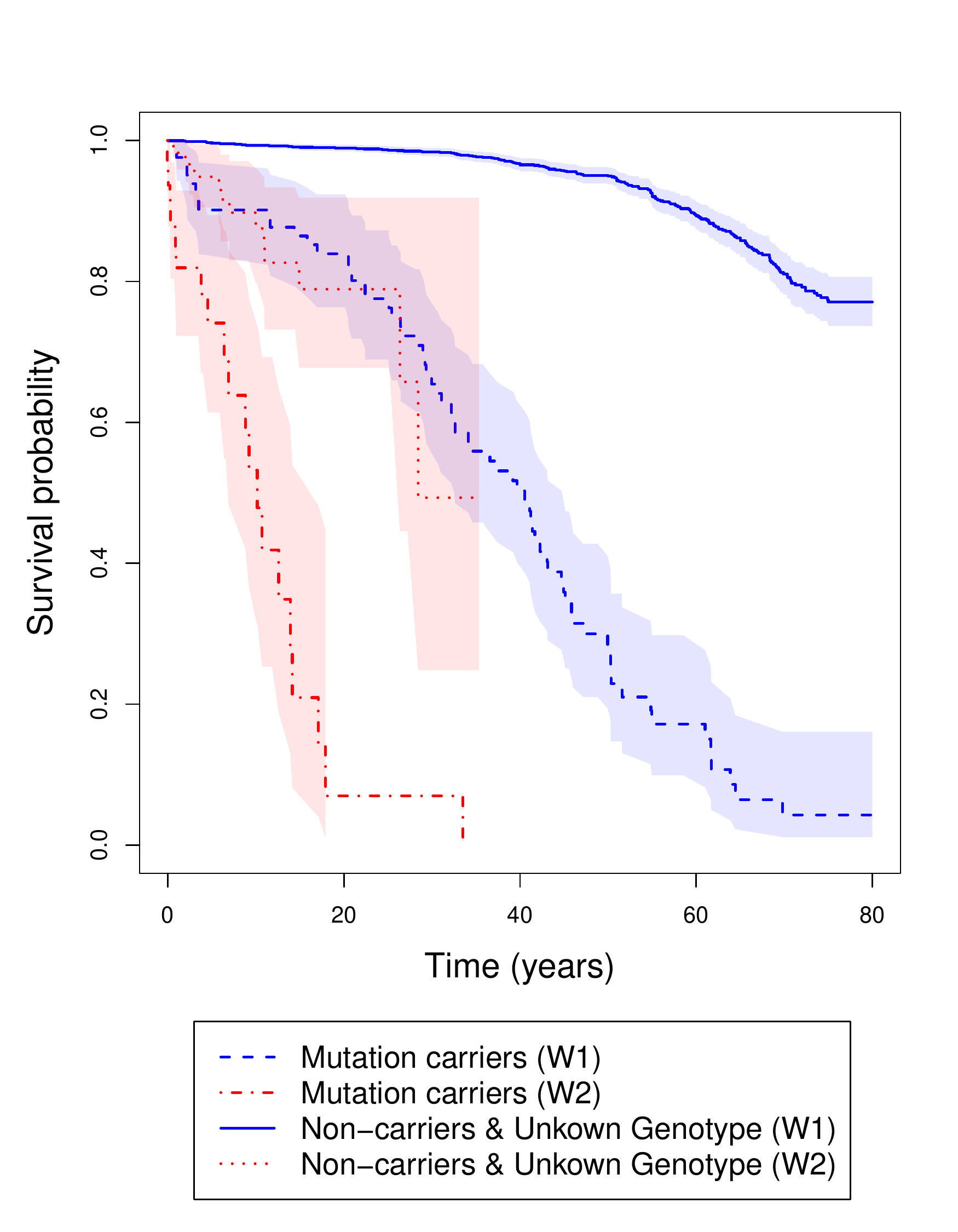}  
}
\caption{Supplementary Figure 9. Kaplan-Meier estimates of the survival distributions for the first or the second gap times of the LFS dataset without probands, referred in Section 2.2. The solid lines denote mutation carriers. The dotted lines denote individuals either with a wildtype or without any genotype information. Blue denotes the first gap time $W_1$ and pink denotes the second gap time $W_2$. The shaded areas are the 95\% confidence bounds. A log-rank test gave p-values $<10^{-7}$ comparing the first and second gap time distributions for individuals that are \textit{TP53} mutation carriers, or otherwise, respectively.} \label{fg::KMestimates}
\end{figure}

\begin{figure} [!htbp]
\centerline{
\includegraphics[width = 0.8\textwidth]{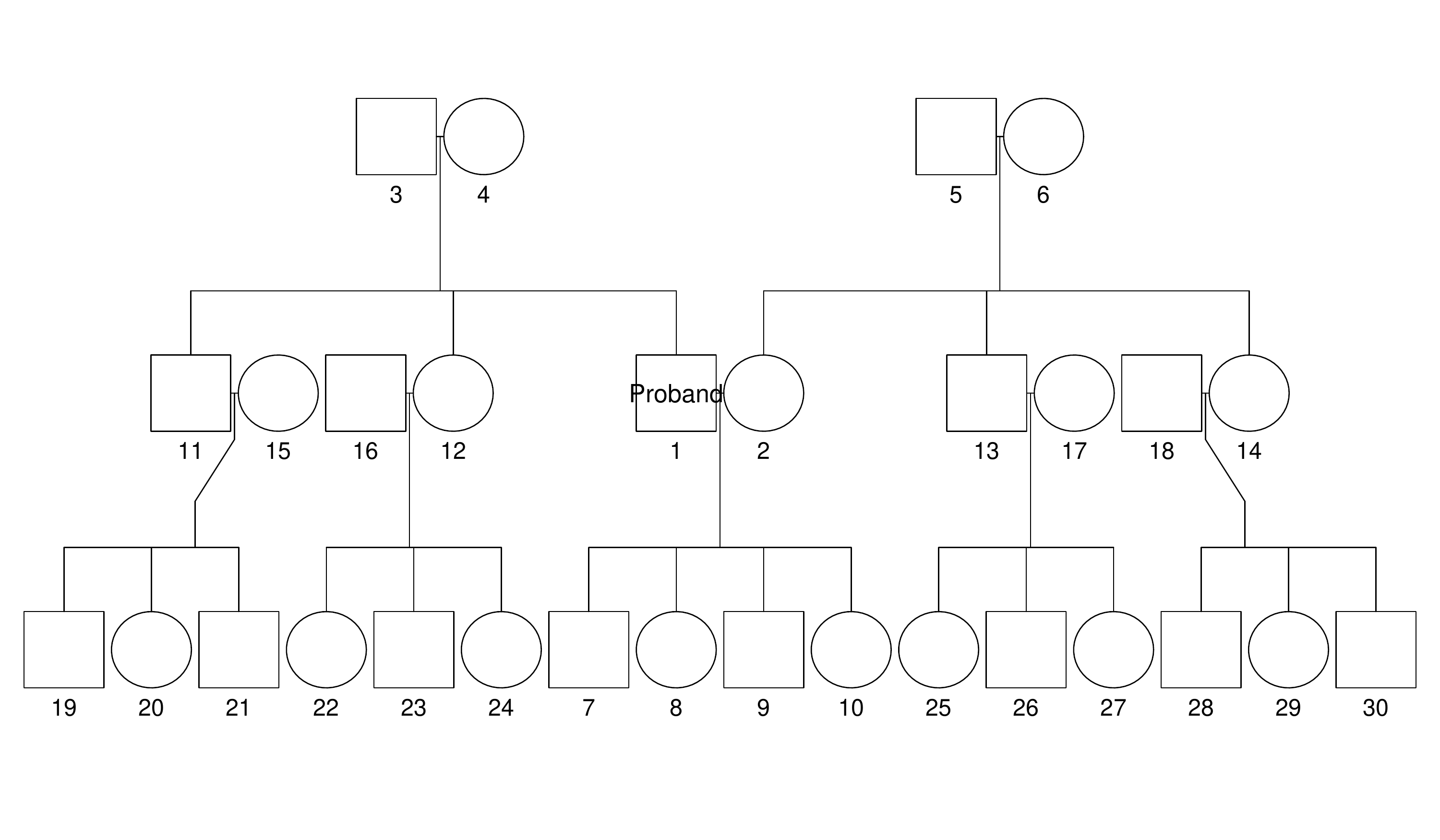}  
}
\caption{Supplementary Figure 10. Illustration of the artificial pedigree structure
    used for the simulation study in Section 6. } \label{fg::sim_ped}
\end{figure}

\newpage 
\begin{table}[!htbp]
\centering
\caption{Supplementary Table 2. Summary of deviance information criterion (DIC) for model selection referred in Section 7.1. *This model is selected.} \label{tb::DIC}
\begin{tabular*}{\columnwidth} {@{}l@{\extracolsep{\fill}} 
                                 l@{\extracolsep{\fill}}
                                 c@{\extracolsep{\fill}}  
                                 r@{}} \hline
Model &           Covariates             								&  DIC  \\ \hline
(M1) & $\{G, S, D(t) \}$    											&  3469.75                 \\
(M2) & $\{G, S, D(t), G\times S \}$    									&  3529.36                 \\
(M3) & $\{G, S, D(t), G \times D(t) \}$    								&  3526.03                 \\
(M4)* & $\{G, S, D(t), G \times S, G \times D(t) \}$ 					&  3478.01                 \\ 
(M5) & $\{G, S, D(t), G \times S, G \times D(t),  S \times D(t) \}$ 	&  3499.61                 \\ 
\hline

\end{tabular*}
\end{table}

\newpage

\bibliographystyle{biorefs}
\bibliography{references}